\documentclass[]{pasj02}

\usepackage[switch,mathlines]{lineno}

\jyear{2025}
\Received{}
\Accepted{}
\def\be{\begin{equation}}\def\ee{\end{equation}}\def\vlsr{v_{\rm LSR}}  
\def\deg{^\circ}  
\def\vrot{V_{\rm rot}} \def\Vrot{\vrot}\def\co{$^{12}$CO }     \def\Tb{T_{\rm B}}       \def\kms{km s$^{-1}$}       
  \def\Tb{T_{\rm B}}    \def\ekms{~{\rm \ km \ s^{-1}}~}  \def\epc{{\rm \ pc} }   \def\apj{ApJ} \def\aap{A\&A} \def\mnras{MNRAS} \def\pasj{PASJ} \def\aj{AJ}           
 \def\Msun{M_{\odot \hskip-4.8pt \bullet}}  
 \def\Osun{\Omega_{\odot \hskip-4.8pt \bullet}}   
\def\Usun{U_{\odot \hskip-4.8pt \bullet}}  \def\Vsun{V_{\odot \hskip-4.8pt \bullet}}    
\def\Wsun{W_{\odot \hskip-4.8pt \bullet}}    
\def\kms{km s$^{-1}$}  \def\deg{^\circ}         
     \def\bc{\begin{center}}\def\ec{\end{center}} 
      
 \def\x{\times}   \def\Rzero{R_0}       
\def\({\left(} \def\){\right)}\def\[{\left[} \def\]{\right]}        \def\ss{\subsection}
  
  \def\sgrastar{Sgr A$^*$}  
   
 \def\G02{G+0.02-0.02+100}  
\def\ekpc{{\rm kpc}}       
 \def\lw{\linewidth}  \def\be{\begin{equation}} \def\ee{\end{equation}}
 \def\nH2{n_{\rm H_2}}  \def\epc{{\rm pc}}  \def\rhomass{\rho_{\rm dyn}} 
\def\Thzero{\Theta_0} \def\Rzero{R_0}  \def\ekmsk{{\rm km~s^{-1} kpc^{-1}}} 
\def\ekpc{{\rm kpc}} \def\vterm{V_{\rm term}}  \def\vterm{V_{\rm term}} \def\vrot{V_{\rm rot}}
\def\sin{\rm sin} \def\be{\begin{equation}} \def\ee{\end{equation}} \def\ss{\subsection}
  \def\Vzero{\Theta_0} 

\begin{document}
\title{The inner rotation curve of the Milky Way} 

\author{Yoshiaki \textsc{Sofue}$^{1}$ and Mikito \textsc{Kohno}$^{2,3}$ } 
\altaffiltext{1}{Institute of Astronomy, The University of Tokyo, 2-21-1 Mitaka, Tokyo 181-8588, Japan } 
\email{sofue@ioa.s.u-tokyo.ac.jp}
\orcid{0000-0002-4268-6499}
\altaffiltext{2}{Curatorial division, Nagoya City Science Museum, 2-17-1 Sakae, Naka-ku, Nagoya, Aichi 460-0008, Japan}
\altaffiltext{3}{Department of Physics, Graduate School of Science, Nagoya University, Furo-cho, Chikusa-ku, Nagoya, Aichi 464-8602, Japan}
\orcid{0000-0003-1487-5417}

\maketitle 

   
\begin{abstract}  
We derived the inner rotation curve (RC) of the Milky Way by applying the terminal velocity method (TVM) to the longitude-velocity diagrams (LVD) made from the large-scale survey data of the Galactic plane in the HI (HI4PI whole sky survey) and \co\ lines (CfA-Chile 1.2-m Galactic plane survey, Nobeyama 45-m Galactic plane and Galactic Center surveys, and Mopra 22-m southern Galactic plane survey).
The derived RC agrees well with the RCs derived from the astrometric measurements of the maser sources by very long baseline interferometer (VLBI) observations {and the GAIA result.
We combined them to construct a unified RC from $R=0$ to $\sim 25$ kpc} and decomposed the curve into bulge, disc and dark halo components with high precision.
The dark matter density near the Sun is estimated to be $0.107 \pm 0.003$ GeV cm$^{-3}$.
We present the RC as ascii tables for the solar constants of $(\Rzero,\Vzero)=(8.178 ~\ekpc, 235.1 \ekms)$,
We also obtained a detailed comparison of the eastern ($l\ge 0\deg$) and western ($< 0\deg$) RCs in the HI and CO lines, which allowed the creation of an E/W asymmetry curve of the velocity difference.
The E/W asymmetry is fitted by a sinusoidal function of the radius with the amplitude increasing toward the Galactic Center.
We consider the possibility of the origin due to a weak bar inside $\sim 4$ kpc. 
\end{abstract}



\section{Introduction} 
\label{intro}

The rotation curve (RC) is the fundamental tool for measuring the dynamical mass and its distribution in the Galaxy under the assumption that the galaxy is axisymmetric \citep{SofueRubin2001,sofue2020}.
Various methods to derive the rotation curve (RC) of the Milky Way have been proposed such as the terminal velocity method (TVM) for the gaseous disk, the radial velocity plus distance method for stars, the disk thickness method for HI disk, the trigonometric method for maser sources and a large number of stars \citep{clemens1985,Fich+1991,1997PASJ...49..453H,SofueRubin2001,Sofue2017,2019ApJ...870L..10M,sofue2020,reid+2019,vera+2020,2019ApJ...871..120E}. 
A large-scale compilation of RC data has been obtained and is available electronically \citep{huang+2016,Iocco+2015,Pato+2017a,Pato+2017b,Krelowski+2018,sofue2020,reid+2019,vera+2020,2019ApJ...871..120E,2023ApJ...942...12W}.

The TVM, which we employ in this work, measures the terminal velocity of the gaseous disk of the Milky Way inside the Solar circle in the HI and CO line emissions under the assumption of axisymmetry
\citep{Burton+1978,clemens1985,alvarez+1990,mcclure+2007,marasco+2017}.
It has the advantage of uniquely determining the galacto-centric distance without suffering from uncertainty in distance measurements.
CO line has the advantage of measuring the motion of molecular clouds sharply concentrated near the Galactic plane with the lowest velocity dispersion among Galactic objects, so that CO traces the rotational kinematics of the Galactic disk most precisely with a smaller influence of random motion and velocity dispersion compared to other species.

This paper has the following structure:
In section \ref{data} we describe the HI and CO data used for the analysis and their parameters. 
In section \ref{method} we describe the method and fundamental assumption on the analysis, and in section \ref{result} we presents the obtained rotation curve.

Although RC is the most powerful tool for estimating the mass of the Galaxy, a debate has long been focused on the influence of the bar on the inner part of the RC. 
Since a bar induces non-circular flow, the RC derived using TVM cannot be transformed to the mass, for which we have to solve the inverse problem from one-dimensional RC to two-dimensional velocity field.
This difficulty is also not solved by simulation, which can treat only a direct problem.
In Section \ref{discussion} we discuss these issues and consider the implications of the derived RC on the Galactic dynamics.

We summarize the related terms:
\begin{itemize}
\item {Longitude-velocity diagram (LVD): Intensity distribution of the radio spectral line (HI or CO line) in the longitude-radial velocity $(l,\vlsr)$ plane. All (radial) velocities are referred to the LSR (Local Standard of Rest) in this paper.} 
\item Terminal velocity ($\vterm$, TV): LSR velocity of a Gaussian component having the maximum center velocity in a line spectrum at a certain longitude in the Galactic plane. If the rotation is circular, which we assume in this paper, this coincides with the tangent velocity. 
TVM stands for the terminal-velocity method, and RC/TVM stands for an RC using TVM.
\item Tangent velocity: LSR velocity of the tangential point of a pure circle around the GC. We do not use this term in this paper.
\item Rotation velocity ($\vrot$): Orbital circular velocity of the LSR around the GC calculated using the terminal velocity by assuming that the Galaxy is rotating circularly (axisymmetric). Equivalent to circular velocity.
\item Rotation curve (RC): A plot of $\vrot$ against the galacto-centric distance $R$.
\item Bar: We assume axisymmetry, allowing for a small perturbation by a weak bar, but a bar-induced fast non-circular motion cannot be discussed using our RC.
\end{itemize}

\section{Data}
\label{data}

\subsection{HI}
 HI survey data were obtained by the full-sky HI4PI survey \citep{2016A&A...594A.116H}, which combine the Effelsberg-Bonn H I Survey (EBHIS; \cite{2016A&A...585A..41W}) and the Galactic All-Sky Survey (GASS; \cite{2015A&A...578A..78K}). 
The observations were performed by the Effelsberg 100-m and Parkes 64-m radio telescopes. The angular and velocity resolution of the combined map is \timeform{16.2'} and 1.49 \kms, respectively. The sensitivity is $\sigma_{\rm rms}\sim 0.043$ K. 
 HI4PI cube data was taken from the CDS website\footnote{https://cdsarc.cds.unistra.fr/ftp/J/A+A/594/A116/}.

\subsection{CO}
We used the $^{12}$CO $J$~=~1--0 line data of the entire inner disc, which was obtained by the 1.2 m radio telescopes in CfA and Chile. These CO data are compiled by \citet{1989ApJS...71..481B},\citet{1997A&AS..125...99B}, and \citet{2001ApJ...547..792D}.
This CO data has an angular resolution of $\sim \timeform{9'}$ and a root mean square (rms) noise level of $\sim 0.1$-$0.3$ K.

More high-resolution $^{12}$CO~$J$~=~1--0 line data in the inner Galactic plane are taken from the Nobeyama 45-m and Mopra 22-m radio telescopes. These CO surveys covered the eastern and western sides of the Galactic disk, which correspond to the first and fourth Galactic quadrants.

The eastern CO surveys are the FOREST unbiased Galactic plane imaging survey with the Nobeyama 45 m telescope (FUGIN\footnote{https://nro-fugin.github.io}:\cite{ume+2017,Torii+2019}) and the Nobeyama 45 m Local spur CO survey \citep{2022PASJ...74...24K,2023PASJ...75..279F}.
The observations were performed by the mutibeam receiver FOREST (FOur beam REceiver System on the 45 m Telescope: \cite{mina+2016,2019PASJ...71S..17N}).
The effective velocity resolution was 1.3 \kms, the typical rms noise of the brightness temperature $\Tb$  was $\sim 1.0$-$1.5$ K, and the effective angular resolution was $\timeform{20"}$, while the original beam of the 45-m telescope at the \co frequency was $\timeform{15"}$. 
The \co line channel maps had a grid spacing of $\timeform{8.5"}\times \timeform{8.5"}\times 0.65 ~{\rm km ~s^{-1}}$ in the $(l,b,\vlsr)$ space.
The FUGIN CO survey data were retrieved from the Japanese Virtual Observatory (JVO) website \footnote{http://jvo.nao.ac.jp/portal/nobeyama/fugin.do}.

The $^{12}$CO $J$~=~1--0 survey data in the Galactic Center (GC) region were obtained by the multibeam receiver of BEARS (25-BEam Array Receiver System: \cite{2000SPIE.4015..237S,2000SPIE.4015..614Y}) installed on the Nobeyama 45 m telescope. 
The angular resolution is $\sim \timeform{15"}$. The grid spacing is $(l,b,V_{\rm LSR})=(\timeform{7.5"},\timeform{7.5"},2\ $\kms$)$.
The typical rms noise level is $\sim 1$ K. 
More detailed data properties of the GC CO survey were described in \citet{2019PASJ...71S..19T}\footnote{https://www.nro.nao.ac.jp/\textasciitilde nro45mrt/html/results/data.html\#GC}.

The western CO survey is the Mopra Southern Galactic Plane CO Survey \citep{2013PASA...30...44B,2015PASA...32...20B,2018PASA...35...29B,2023PASA...40...47C}\footnote{https://mopracosurvey.wordpress.com}.  The survey data are converted to the main beam temperature ($T_{\rm MB}$) from the antenna temperature ($T_A^*$) using the relation of $T_{\rm MB} = T_{\rm A}^*/\eta$ with the extended beam efficiency of $\eta = 0.55$ \citep{2005PASA...22...62L}. The spatial and velocity resolutions of the data release 4 (DR4) are $\timeform{36"}$ and 0.1 \kms, respectively. The grid spacing of the cube data using this paper is $(l,b,V_{\rm LSR})=(\timeform{30"},\timeform{30"},1\ $\kms$)$.
The typical rms noise level is $\sim 0.7$ K.
The Mopra CO survey data were retrieved from the portal website\footnote{https://doi.org/10.25919/9z4p-mj92}.

\section{Terminal-velocity method (TVM) to determine rotation curve}
\label{method}
\def\lw{\linewidth}

\begin{figure*}[t]
\begin{center}
\includegraphics[width=\lw]{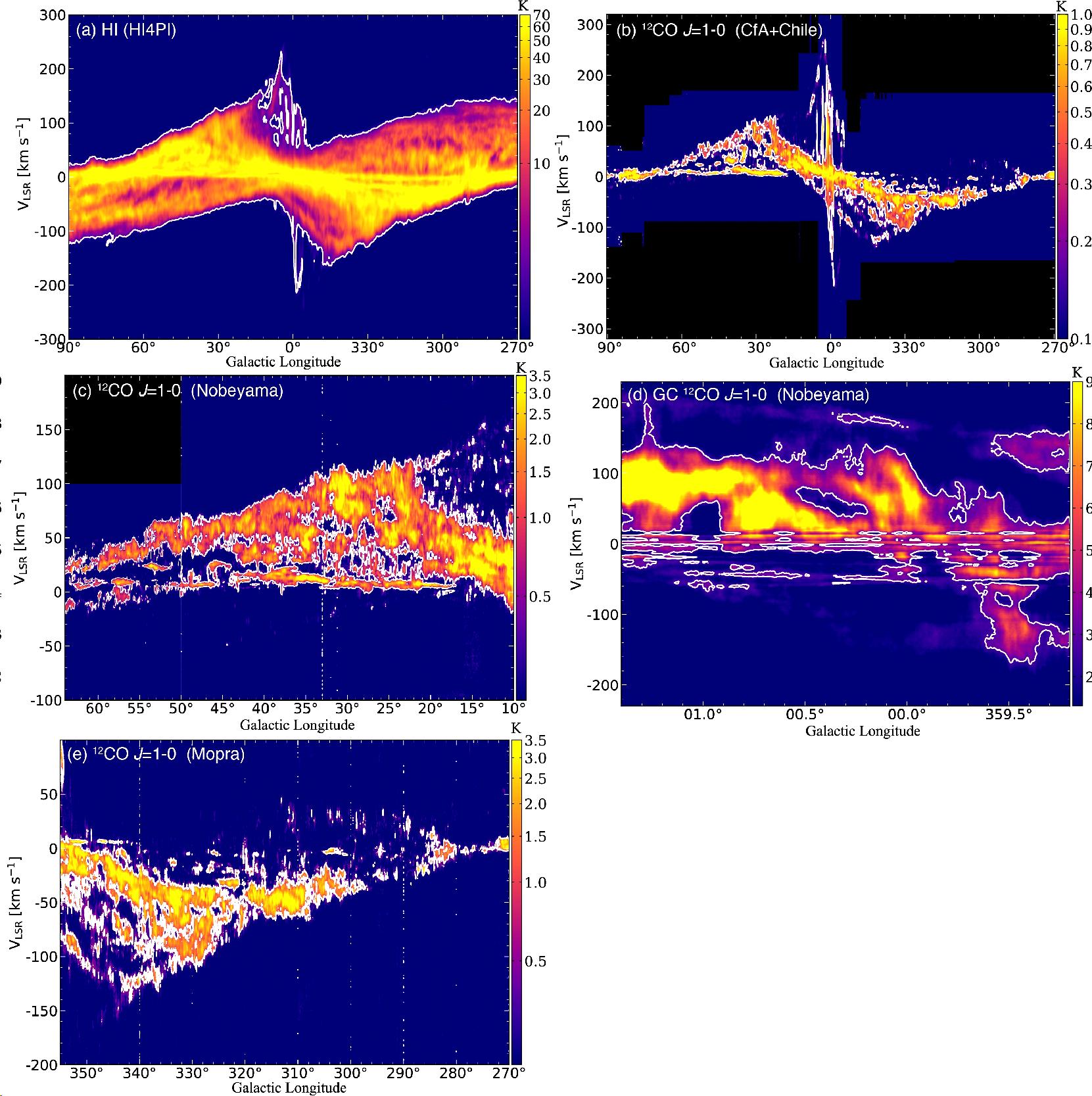}
\end{center}
\caption{HI and CO-line LVDs of the inner Galaxy. (a) HI 21 cm taken from the Effelsberg 100-m and Parkes 64-m radio telescopes \citep{2016A&A...594A.116H}. (b) $^{12}$CO $J$~=~1--0 obtained by CfA and Chile 1.2-m telescopes \citep{2001ApJ...547..792D}. (c) $^{12}$CO $J$~=~1--0 obtained by the Nobeyama 45-m telescope with the FUGIN and Local spur CO survey project \citep{ume+2017,Torii+2019,2022PASJ...74...24K,2023PASJ...75..279F}. (d) $^{12}$CO $J$~=~1--0 obtained by the Nobeyama 45-m telescope with the GC CO survey \citep{2019PASJ...71S..19T}. (e) $^{12}$CO $J$~=~1--0 taken from the Mopra Southern Galactic Plane CO Survey \citep{2023PASA...40...47C}. The white contour levels are {2, 0.25, 0.5, 1.2, and 2.8 K} from (a) to (e), respectively.
{Alt text: HI and CO line longitude-velocity diagrams from the large-scale Galactic plane surveys.}
} 
\label{fig1} 
\end{figure*} 

\ss{Galactic constants}

The solar angular velocity about the GC has been observed by parallactic measurements of maser sources using VLBI Exploration of Radio Astrometry (VERA:\cite{vera+2020})
\be \Osun=(\Thzero+\Vsun)/\Rzero=30.17\pm0.3  ~\ekmsk \ee 
and Very Long Baseline Array (VLBA:\cite{reid+2019}):
\be\Osun=30.32\pm 0.27~\ekmsk. \ee
In this paper, we adopt the mean of these two measurements:
\be \Osun=30.25\pm 0.4 ~\ekmsk,\ee 

The peculiar velocity of the Sun corresponding to the local standard of rest (LSR) has been measured by various methods (Table 1 in \cite{Huang+15}).
The solar local motion is assumed to have the following values \citep{sch+2010}:
\be(\Usun,\Vsun,\Wsun)=(11.1^{+0.69}_{-0.75},12.24^{+0.47}_{-0.47},7.25^{+0.37}_{-0.36})\ekms, \ee
with
\be \sqrt{\Usun^2+\Vsun^2+\Wsun^2}=18.04 ~\ekms. \ee

The Galactic constant, $\Rzero$, or the distance to \sgrastar\ is measured by analyzing the orbital parameters of high-speed stars orbiting close to the nucleus \citep{gravity+2019}:
\be \Rzero=R_{\rm Sgr~A^*}= 8.178 \pm 0.013_{\rm sta} \pm 0.022_{\rm sys} ~\ekpc \ee

Adopting this distance and the average of the VERA and VLBA results for solar angular speed, 
we obtain the solar circular velocity as 
\be
\Thzero+\Vsun=\Osun \Rzero= 247.3\pm 2.5 ~\ekms.
\ee
Correcting for the azimuthal solar proper motion $\Vsun$, we obtain another Galactic constant $\Thzero$, or the circular velocity of the LSR around \sgrastar\ as
\be
\Thzero=(247.3\pm2.5) - (12.2\pm 0.5)=235.1\pm2.5 ~\ekms:
\ee

To summarize, in this paper we assume the following Galactic constants:
\be (\Rzero,\Thzero)\simeq (8.18 ~\ekpc, ~235.1 ~\ekms). \ee

\ss{Rotation velocity from terminal velocity }

Figures \ref{fig1} (a-e) show the LVDs (brightness temperature $\Tb$) in the HI and CO line emissions against longitude in the Galactic plane. 
{In this paper, we used the data at $|b| < \timeform{3D}$ in the HI4PI and CfA-Chile CO surveys. 
In the Nobeyama and Mopra CO surveys, we used data at $|b| < \timeform{1D}$.
In the GC survey, we used the data at $|b| < \timeform{0.3D}$ due to the limited observation range of latitude. We applied TVM to these LVDs}

The Galacto-centric distance of the tangent point, which is assumed to coincide with the position having the terminal velocity under the axisymmetric assumption, is defined by
\be
{R=\Rzero ~\sin}~l.
\ee
The rotation velocity, $\vrot (R)$, of this point is calculated using the terminal LSR velocity $\vterm$ by
\be
{\vrot(R)=\vterm + \Thzero~ \sin}~l.
\label{eq_rc}
\ee 
Note that the values become negative at $l=180\deg$ to $360\deg$.
The rotation curve is defined as a plot of $|\vrot|$ against $|R|$, but is usually presented without notice.

\begin{figure} 
\begin{center}    
\includegraphics[width=\lw]{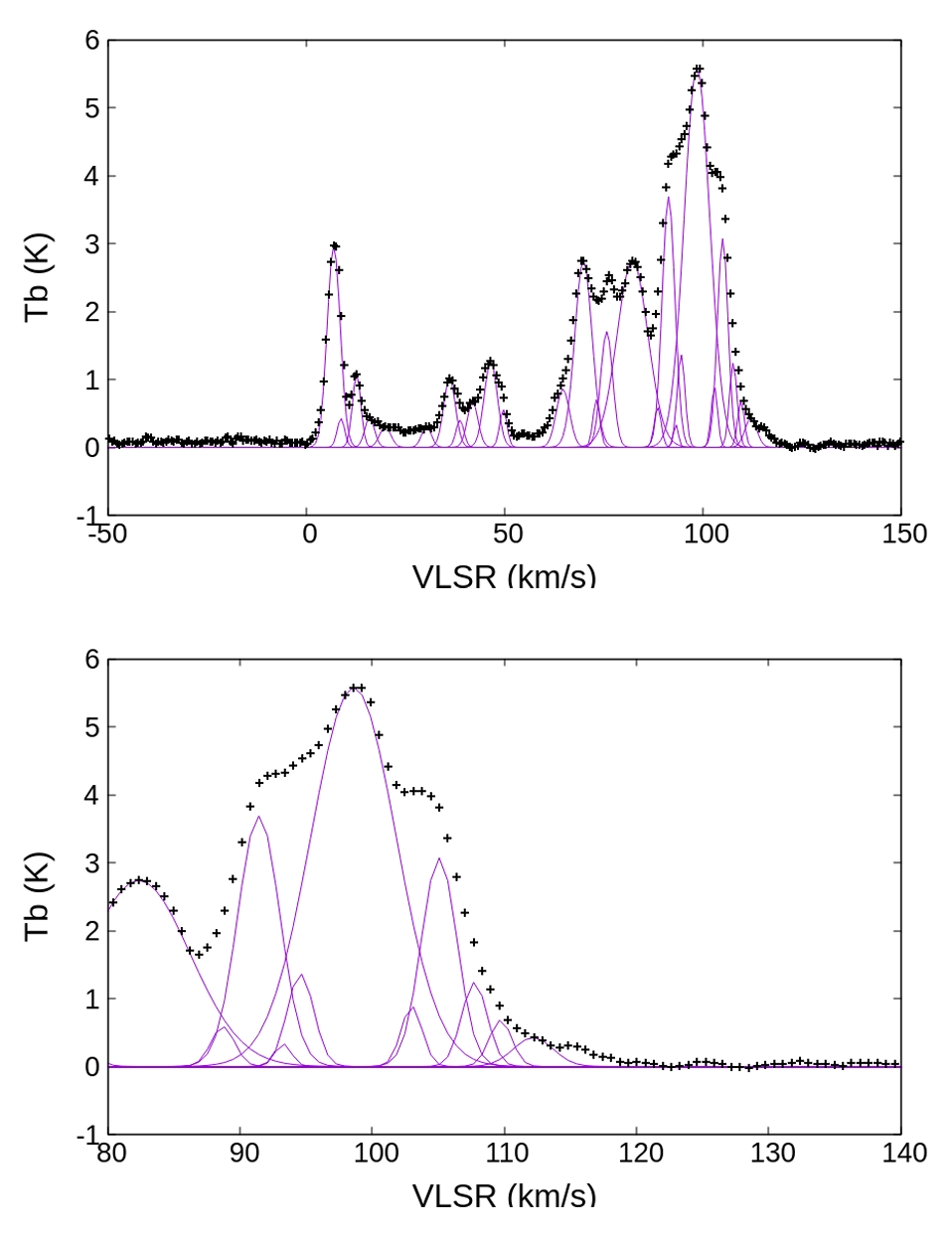} 
\end{center} 
\caption{Line profile at $l=30\deg$ decomposed into Gaussian components.
{Alt text: Diagram explaining Gaussian deconvolution of a line spectrum.}
} 
\label{fig2}
\end{figure}

The terminal velocity $\vterm$ is determined by the following method.
The simplest way to determine the terminal velocity is to pick up the highest-velocity component after deconvolution of the line profile into many components.
Figure \ref{fig2} shows the CO line spectra in the Galactic plane at several longitudes around $l=31\deg$.
Each spectrum can be expressed by the superposition of many components, each represented by a Gaussian profile, as indicated by the red lines.   
    
We apply the Gaussian decomposition to each spectrum of the CO line profile of the FUGIN survey. 
Since the terminal velocities read from the data at higher latitudes tend to lead to a lower-velocity rotation curve, here we measure the terminal velocities using LVDs {averaged at $|b| < \timeform{1D}$.}
Figures \ref{fig2} show Gaussian running-averaged plots of circular velocities, or the rotation curves, where both the radius interval and the Gaussian half-width were taken to be 20 pc. 

Using the LV plot in figure \ref{fig1} we thus obtain a plot of maximum terminal velocities against longitude, which we adopt as circular velocities.
In order to avoid largely deviating data points from the main LV ridge of terminal velocities, we removed data points that exceed $\pm 30$ \kms from the expected mean value. 

\ss{Correction for line width due to local velocity dispersions}

The HI and CO lines are broadened by the internal velocity dispersion as a result of the turbulent motion of the ISM.
This results in a larger terminal velocity at the solar position, or the $|\vterm|$ is greater than zero.
We correct for the intrinsic velocity dispersion using the following relation:
\be\vterm=V_{\rm term, measured} \pm \sigma_v \ee
with $\pm$ corresponding to negative and positive longitudes, respectively.
We provisionally assume certain values of correction for the HI and CO lines, respectively, and the obtained rotation curve is used to determine their values iteratively so that the smoothed curve meets the solar value of $\Thzero=235.1$ \kms at $\Rzero=8.178$ kpc.
We thus obtained $\sigma_v=15$\kms for HI and 5 \kms for CO, which have been used in the following results.

\section{Result}
\label{result}

\begin{figure}
\begin{center}    
\includegraphics[width=\lw]{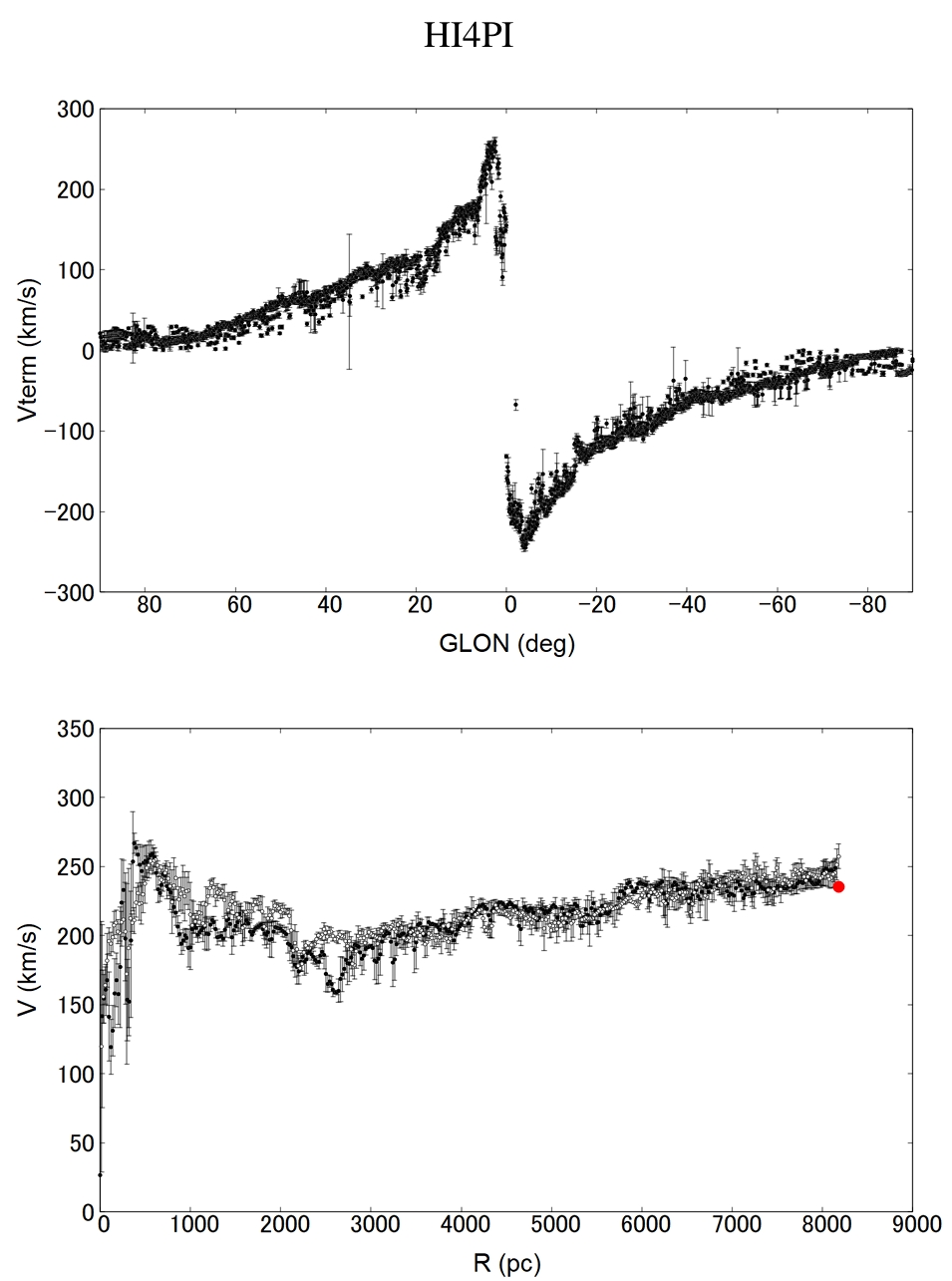}   
\end{center} 
\caption{[Top] HI terminal velocity $\vterm$ after correction (subtraction) of an intrinsic velocity dispersion of $\delta V_{\rm HI}=15$ \kms.
[Bottom] RC East (dot) and RC West (circle). Large dot is the Sun.
{Alt text: HI-line terminal velocity and rotation curve.}
} 
\label{hi}
\end{figure}

\begin{figure}
\begin{center}    
\includegraphics[width=\lw]{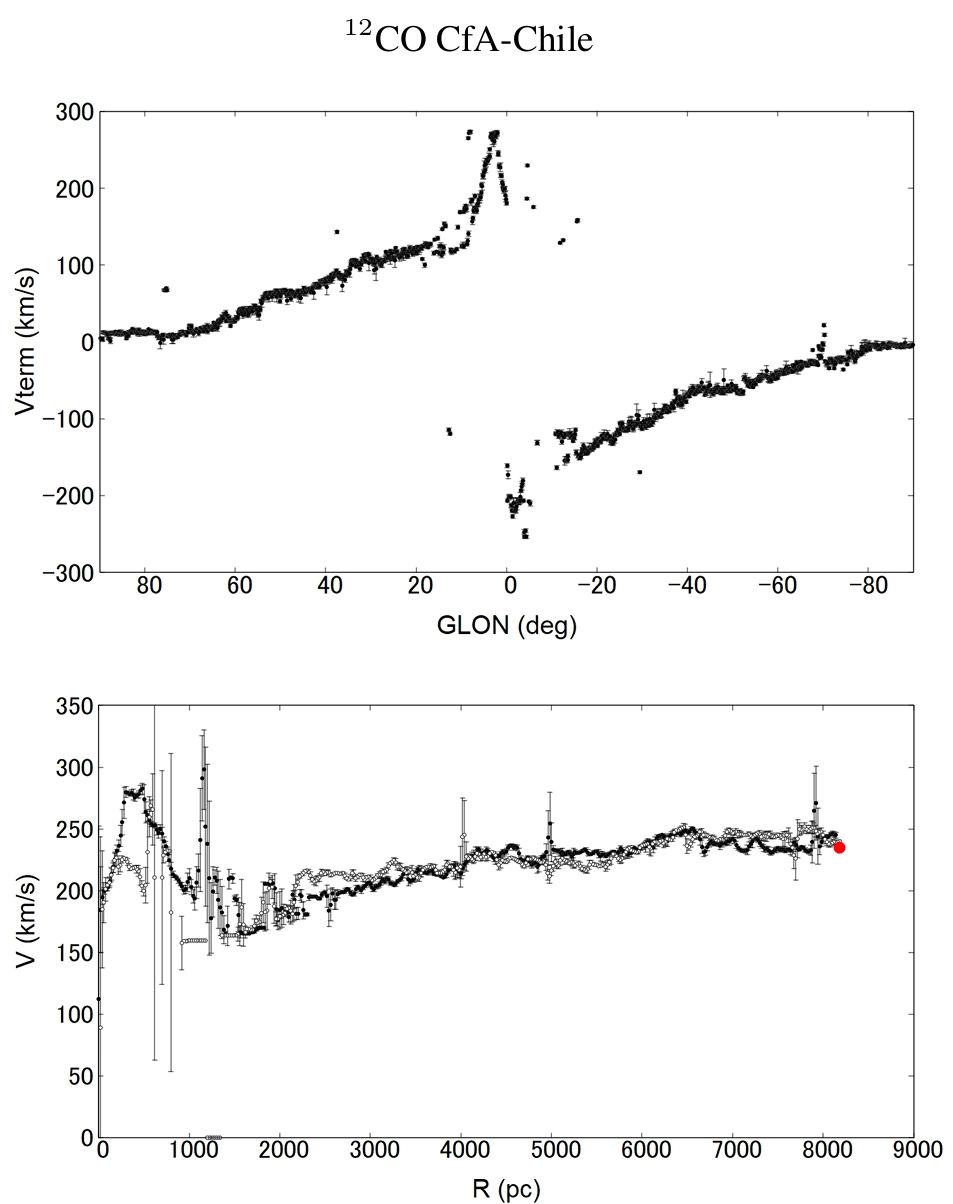}  
\end{center} 
\caption{[Top] \co-line $\vterm$ from CfA-Chile survey. 
 [Bottom] Rotation curves in the E (dots) and W (circles) sides of GC. Big dot is the Sun. 
 {Alt text: Terminal velocity and rotation curve from CfA-Chile CO survey.} } 
\label{cfa}
\end{figure}

\begin{figure} 
\begin{center}   
\includegraphics[width=\lw]{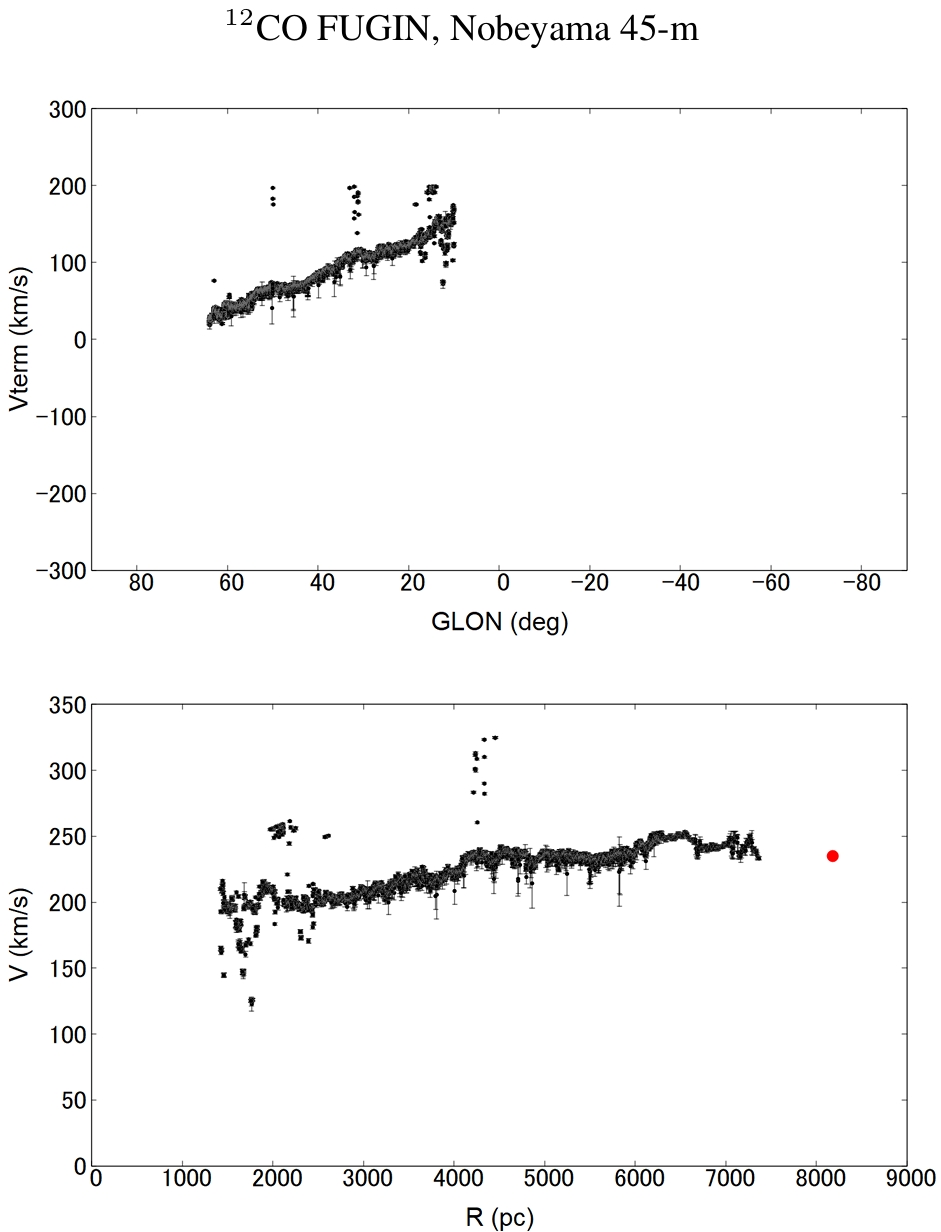}  
\end{center} 
\caption{
[Top] CO FUGIN Terminal velocity against longitude.
 [Bottom] Rotation velocity against radius.  
 {Alt text: Terminal velocity and rotation curve from FUGIN CO-line survey.}} 
\label{fugin}
\end{figure}

\begin{figure} 
\begin{center}   
\includegraphics[width=\lw]{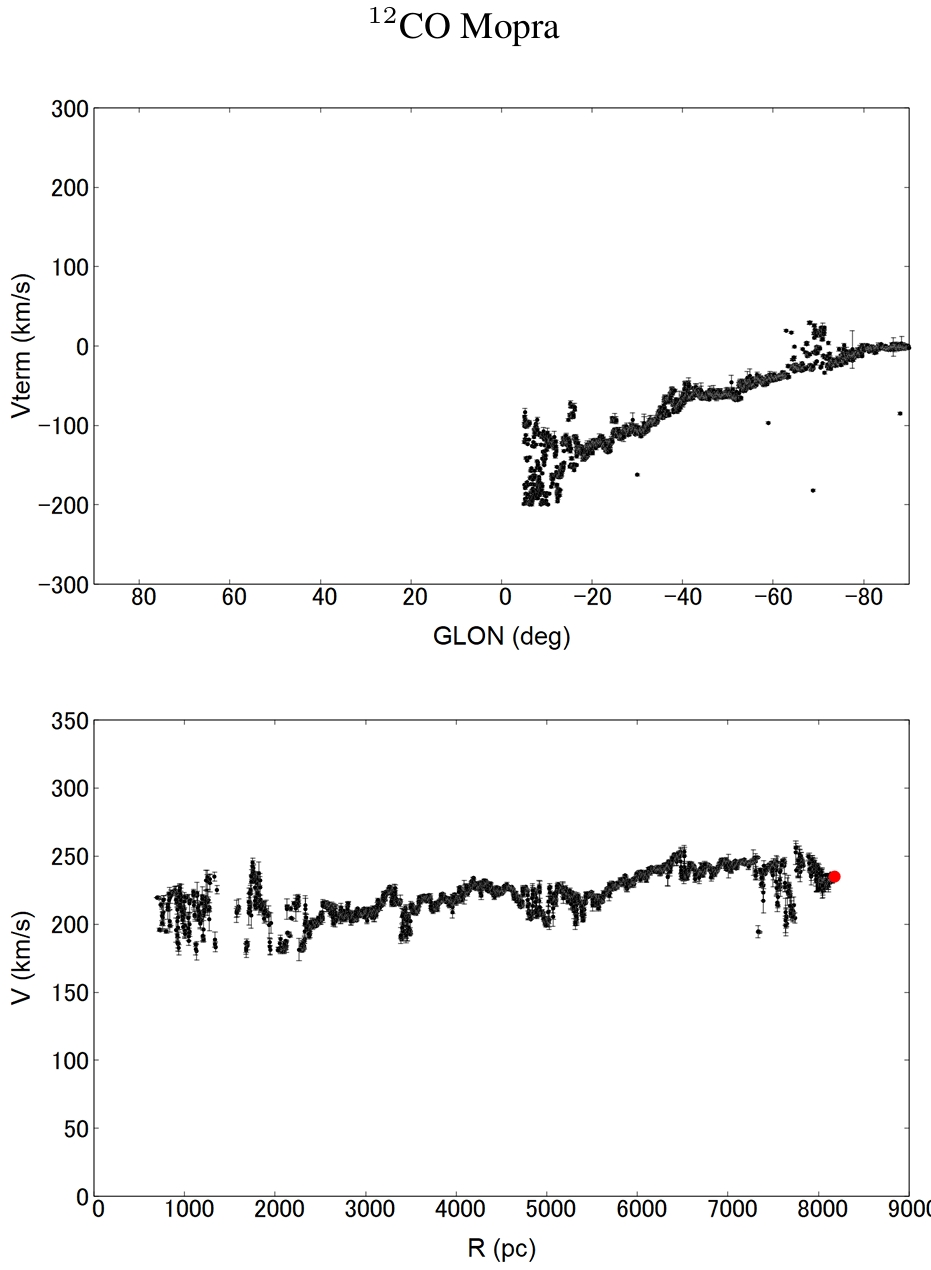}  
\end{center} 
\caption{[Top] Mopra CO terminal velocity. 
[Bottom] Rotation curve.
{Alt text: Terminal velocity and RC from Mopra CO line survey}} 
\label{mop}
\end{figure}

\begin{figure}
\begin{center}    
\includegraphics[width=\lw]{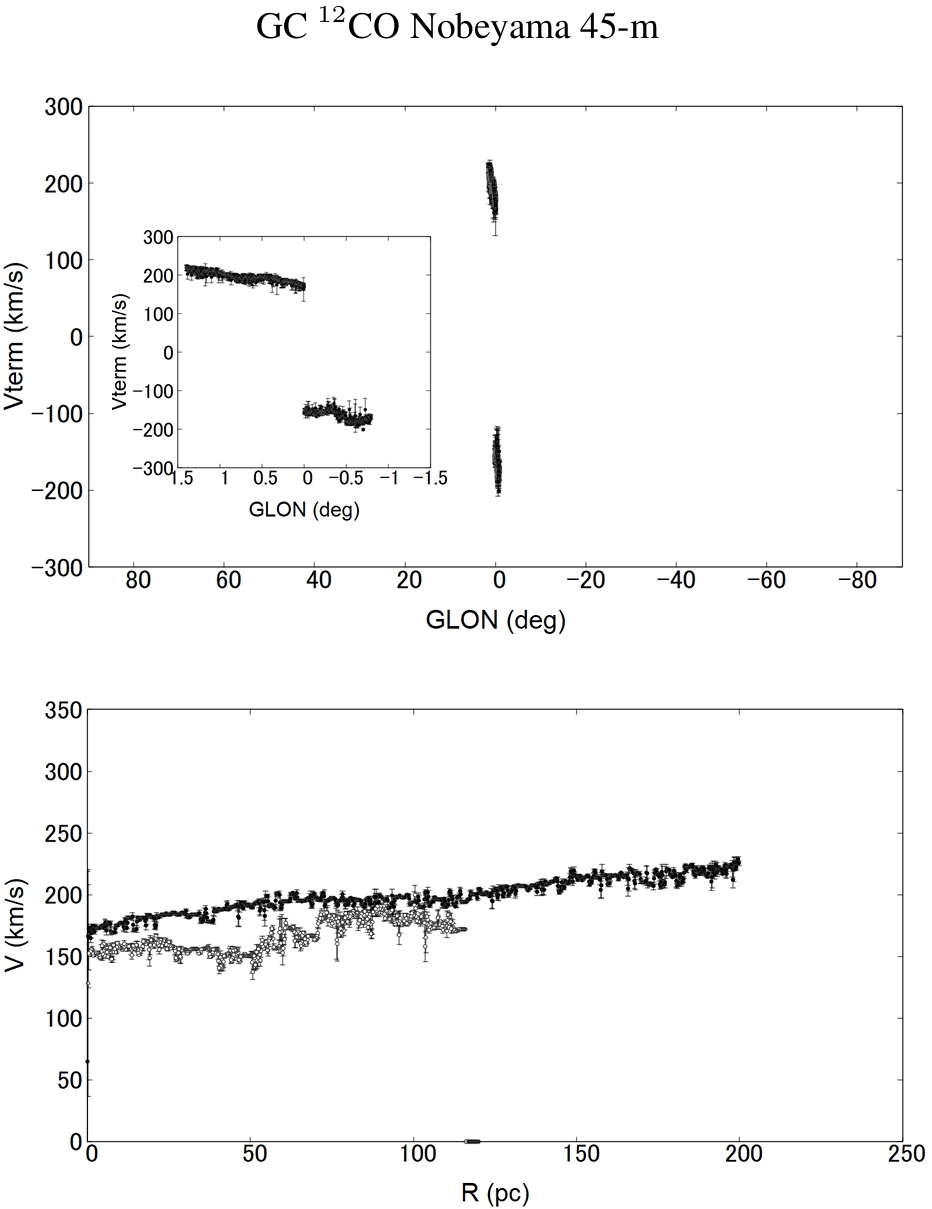}       
\end{center} 
\caption{[Top] GC \co-line terminal velocity from LVD averaged between $-0\deg.3 \le b \le 0\deg.3$. 
The threshold intensity for deconvolution was taken at 5 K in order to avoid contamination by the expanding molecular ring (EMR) at high velocities. 
[Bottom] GC CO line $\vrot$.
{Alt text: Terminal velocity and rotation curve of the GC in the CO line.}
} 
\label{gc}
\end{figure}

\ss{Terminal velocity and rotation curves}
In figures \ref{hi} to \ref{gc} we show the terminal velocities obtained and the corresponding rotation curves.
The top panels show $\vterm$ plotted against longitude, and the bottom panels present rotation curves plotted against the galacto-centric distance $R$.

Figure \ref{hi} shows the result using the HI LVD averaged from $b=-3\deg$ to $+3\deg$ using the HI4PI survey.
The bottom panel shows the rotation curves separately for the eastern (dots) and western (circles) sides of the GC.
Figure \ref{cfa} shows the result using an LVD averaged between $b=-3\deg$ and $+3\deg$ from the CfA-Chile \co\ line survey of the Galactic plane. 
Figure \ref{fugin} is the result using the \co\ LVD averaged between $b=-1\deg$ and $+1\deg$ of the FUGIN galactic plane surveys. 
Figure \ref{mop} is the result using \co\ LVD averaged between $b=-1\deg$ and $+1\deg$ from Mopra.
Figure \ref{gc} is the result of the GC survey on the line \co {averaged between $b=-0.3\deg$ and $+0.3\deg$} using the Nobeyama 45-m telescope.

The rotation velocity was calculated as follows.
\be
\vrot(R)={\Sigma_1^N ~\vrot(R_i) ~w_i \over \Sigma_1^N ~w_i}, 
\ee
where $N$ is the total number of data points (measurements), $\Vrot(R_i)$ is the derived value of $\vrot$ for the $i$-th data point at radius $R_i$, and $w_i$ is the weight given by
\be
w_i=\exp~ \left[ -\left({R_i-R \over \delta R}\right)  ^2 \right]
\ee
with $\delta R$ (=2 pc here) being the half width for the Gaussian averaging.
The error is calculated by
\be
\delta V=\sqrt{\Sigma [\vrot(R_i)-\vrot(R)]^2 w_i \over \Sigma ~ w_i}.
\ee

\ss{The inner RC of the Milky Way}

\begin{figure*}
\begin{center}      
\includegraphics[width=0.7\lw]{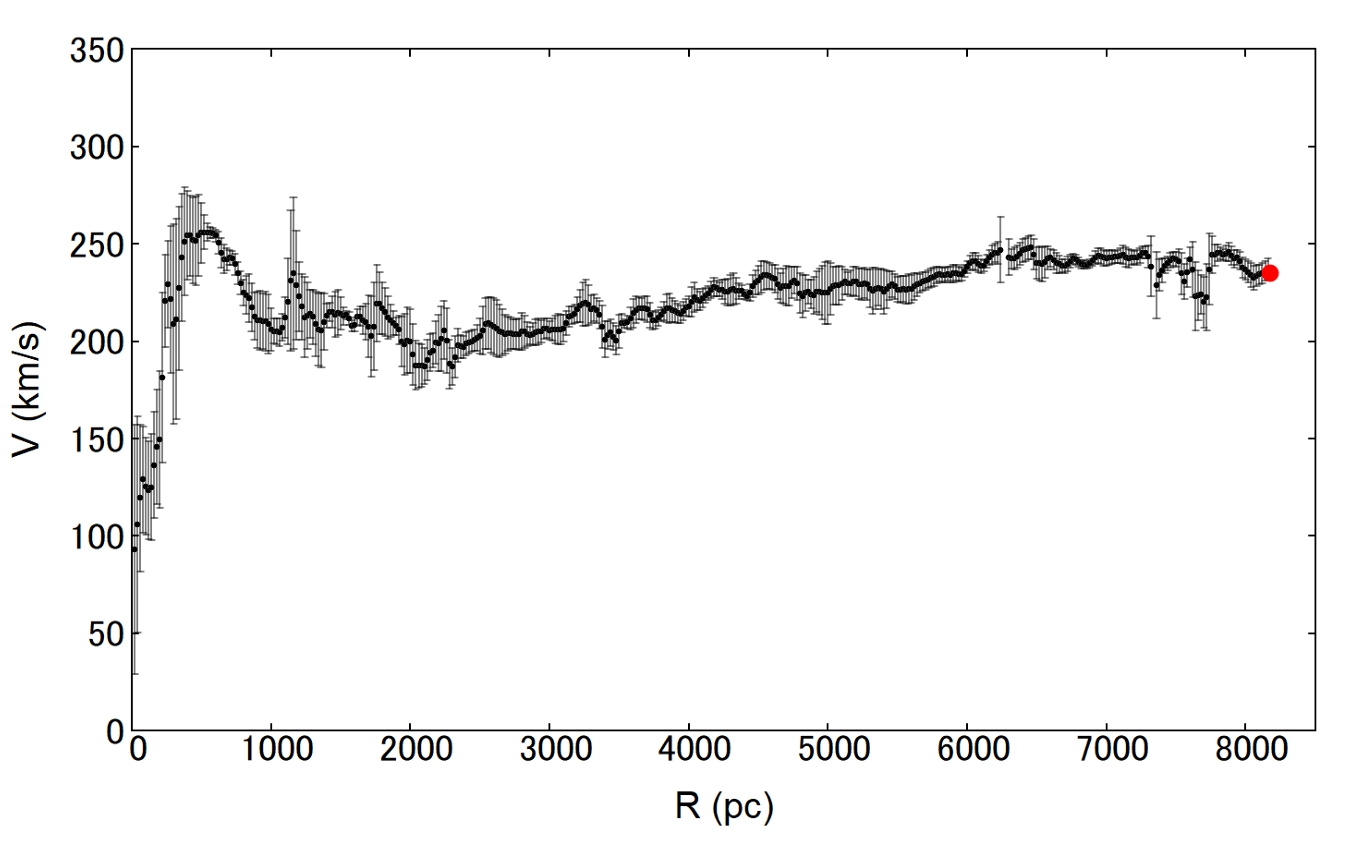}  
\end{center}   
\caption{The inner RC of the Milky Way using HI+CO-line terminal velocities with a radial bin size $\delta R=20$ pc. The big dot is the Sun. 
{Alt text: Rotation curve of the Milky Way inside the solar circle.} 
} 
\label{rc} 

\begin{center}       
\includegraphics[width=0.7\lw]{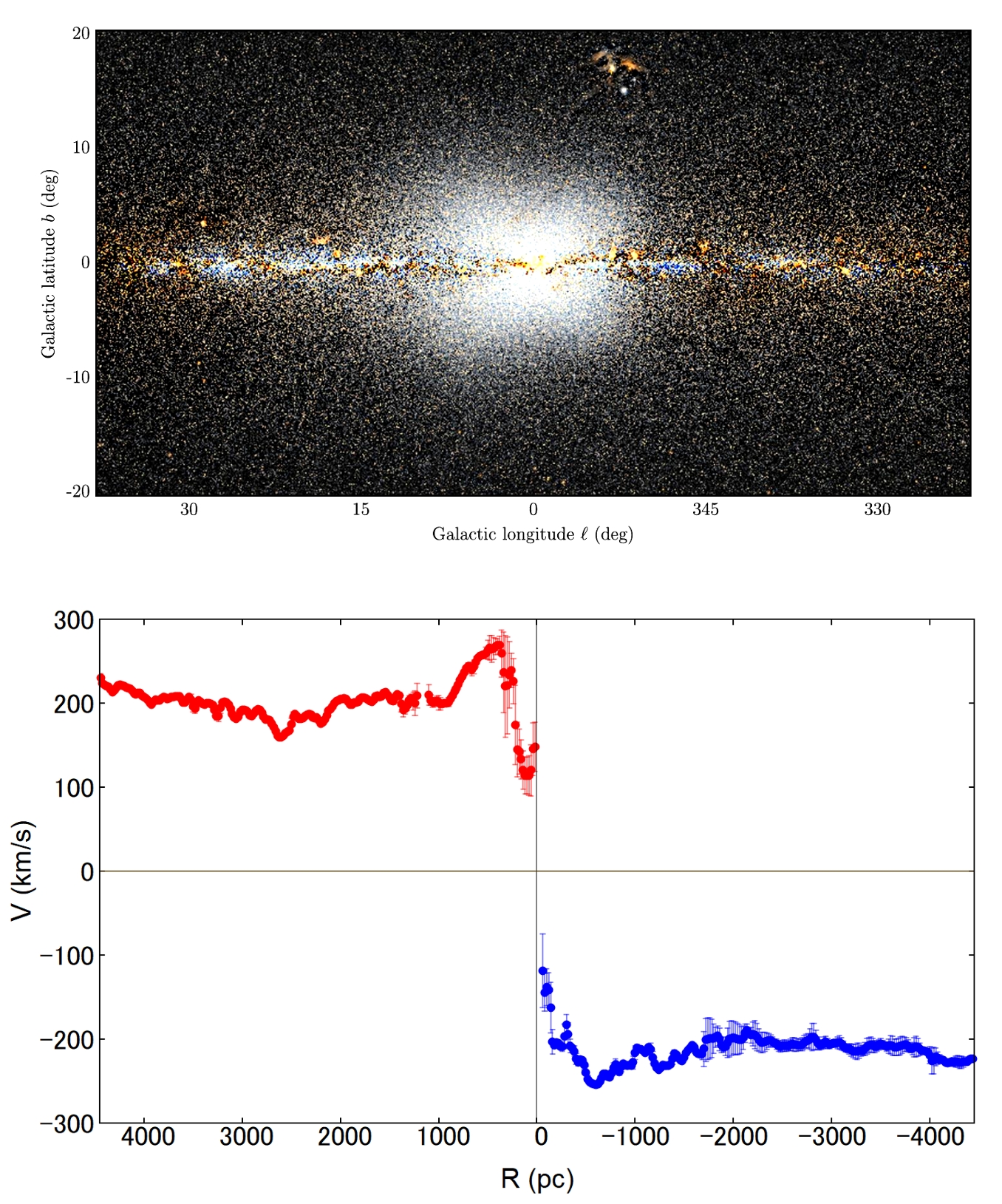}   
\end{center}   
\caption{{Inner E/W RC of the Milky Way compared with the 2.2 $\mu$m image of the galactic bulge \citep{2016AJ....152...14N} (Courtesy: unWISE / NASA/JPL-Caltech / Dr. D. Lang (Perimeter Institute) \& Dr. M. K. Ness (Columbia University)).
Note the synchronous east-west asymmetry: the east side has a larger photometric luminosity and a faster maximum rotation than the west side. {Alt text: Infrared image of the Galactic bulge compared with the E/W Rotation curve.} }
} 
\label{bulge} 
\end{figure*}

Using all the obtained rotation velocities in the HI and CO lines, we construct a rotation curve as shown in figure \ref{rc}, where $\vrot$ values are plotted every 2 pc.
Each dot is a mean of the neighboring points after Gaussian averaging all the data presented in the previous subsection with a half width of $\delta R=2$ pc.
The vertical bars present the standard deviation of the averaged data.

{In figure \ref{bulge} we compare the E/W RCs with the 2.2$\mu$m image of the Galactic bulge \citep{2016AJ....152...14N}.
This figure demonstrates the relationship of the high-velocity peak of the rotation curve at $R\sim 0.3-0.5$ kpc with the massive central bulge.
It is interesting to note the synchronized E/W asymmetry between the photograph and RC: the larger (more luminous) eastern bulge is associated with a faster rotation peak than in the west.}

We plot the HI + CO RC together with the VLBA \citep{reid+2019}, VERA \citep{vera+2020}, and GAIA \citep{2019ApJ...871..120E} results in figure \ref{compare}. 
In the appendix, we show an RC binned at 50 pc and list a table of the plotted values together with a unified RC combined with other observations for wider area. 

\section{Discussion}
\label{discussion}

\ss{Dynamical mass distribution}
\label{ss_mass}

We estimate the approximate mass distribution in the Galaxy assuming spherical symmetry.
The spherical assumption gives a reasonable approximation to the bulge and the halo, while it underestimates the mass of the disc by $\sim 10$ percent compared to the value calculated for a thin disc \citep{sofue2020}.
The nonaxisymmetric effects by the bar and spiral arms will be discussed in the following sections.

\begin{figure}
\begin{center}      
\includegraphics[width=\lw]{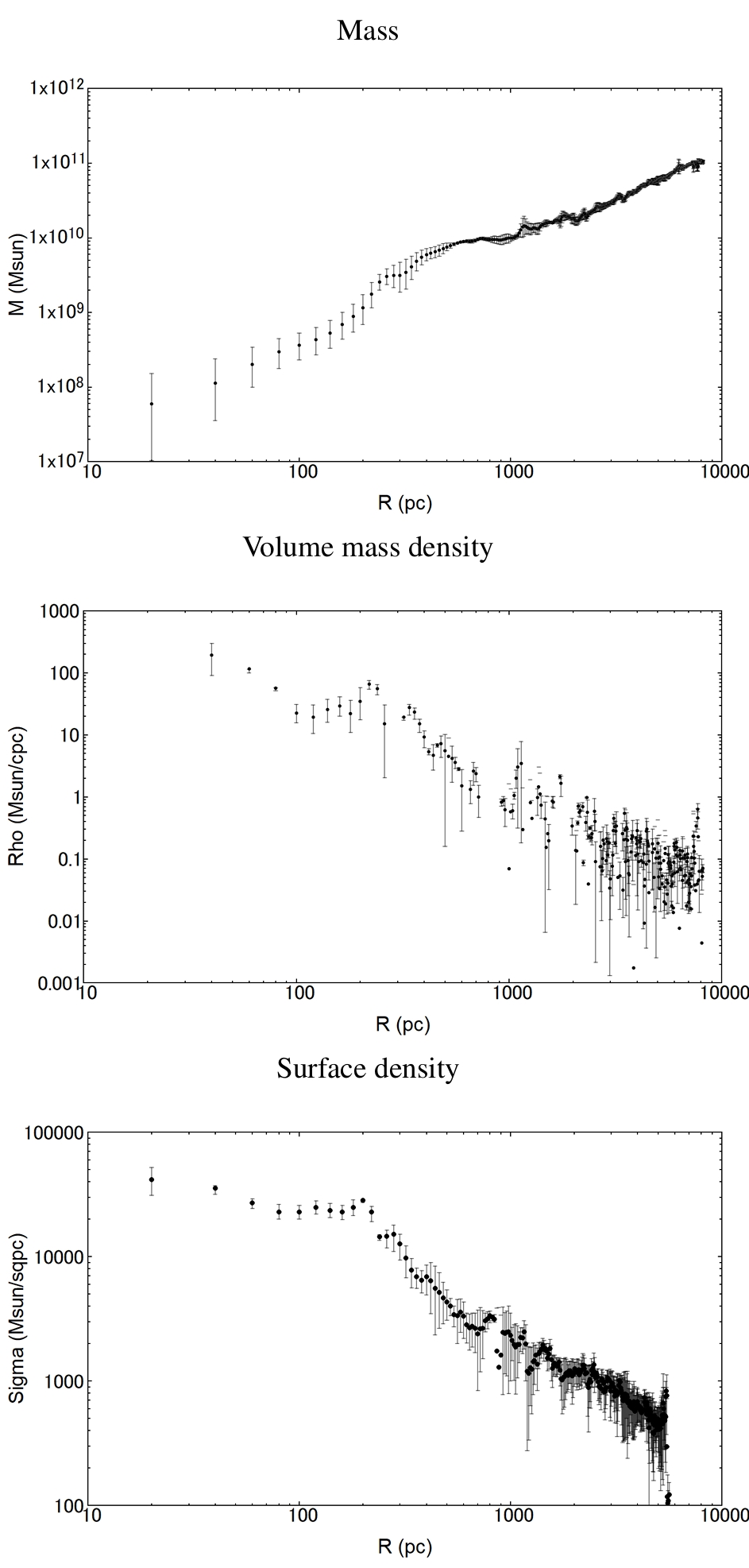} 
\end{center} 
\caption{
[Top] Dynamical mass inside $R$ by spherical assumption. 
[2nd] Mass density by spherical assumption.
[3rd] Surface density distribution using equation \ref{eq_sigA}.  
{Alt text: Radial distributions of the dynamical mass, volume density, and surface density. }
} 
\label{mass}
\end{figure}

The dynamical mass enclosed in radius $R$ is given by
\be
M(R)={R\vrot^2 \over G}.
\label{eq_mass}
\ee 
The volume density is calculated by
\be
\rhomass (R) = {1 \over 4 \pi R^2} {dM(R)\over dR}.
\label{eq_rho}
\ee
The surface mass density (SMD) is calculated by 
\be
\Sigma(R) = 2\int_0^\infty \rhomass(\sqrt{x^2+R^2}) dx, 
\label{eq_sigA}.
\ee
The mass, volume density, and surface mass density obtained here are plotted in figure \ref{mass} as a function of the radius $R$.

\ss{Comparison with outer RCs}
\label{rccomparison}

\begin{figure*}
\begin{center}      
\includegraphics[width=0.7\lw]{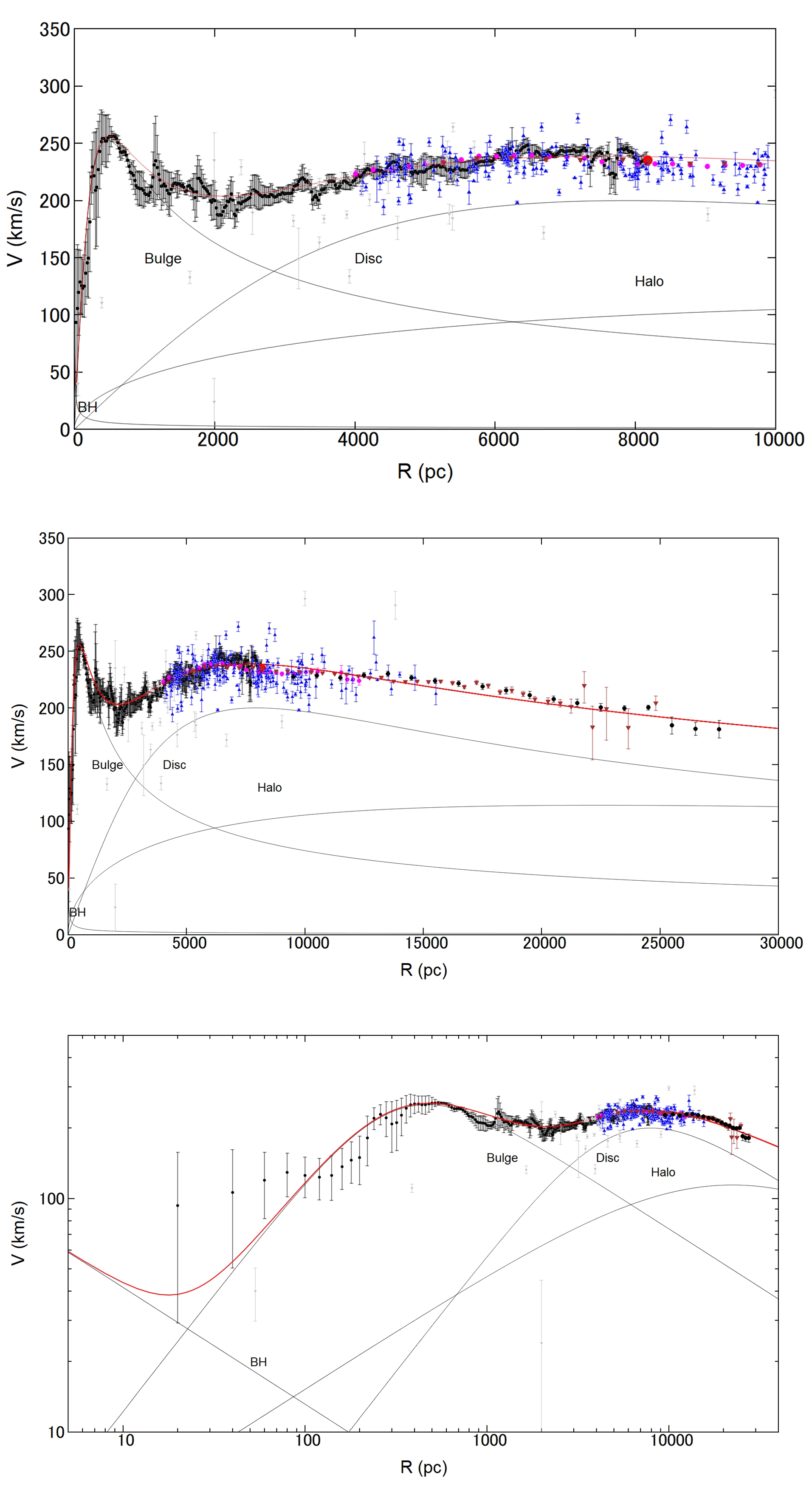}
\end{center} 
\caption{Comparison of the gas (HI+CO) RC by TVM with the results by VERA (triangles, scaled to $\Rzero,\Vzero$) \citep{vera+2020}, VLBA (magenta circles) \citep{reid+2019} and {Gaia (inverse triangles, dots at $\ge 9$ kpc) \citep{2019ApJ...871..120E,2023ApJ...942...12W}.}
Model RCs are shown by the lines with SMBH + Bulge (plummer) + Disc (plummer) + DH (NFW) and their sum.  
{Alt text: Combined plots of the inner RC obtained in this paper with other observations together with a model fitting by decomposition of the mass components.}
} 
\label{compare}
\end{figure*}

In figure \ref{compare} we plot the present RC together with the results by astrometric observations with VERA \citep{vera+2020}, VLBA \citep{reid+2019},and 
{Gaia \citep{2019ApJ...871..120E,2023ApJ...942...12W,2023A&A...678A.208J}}.
The velocities from the literature are corrected for the small differences from our $\Thzero=235$ \kms\ linearly, which are $+4$, $+1$ (not corrected here) and $+6$ \kms\ for VERA ($\Vzero=239$ \kms\ for their $\Rzero$), VLBA (236 \kms) and Gaia (229 \kms) values, respectively.
The higher-order correction due to non-linearity is neglected. 
The radius correction due to slightly different $R_0$ is not applied.

We find that, at $R\gtrsim 2$ kpc, our RC using TVM (RC/TVM) is in good agreement with the RCs from the astrometric measurements of the 3D motions and parallactic distances of a number of maser sources that are distributed over the Galactic disc. 
However, inside $\lesssim 2$ kpc, the rotation velocities of the maser sources are largely scattered because of some 'outliers' \citep{reid+2019,vera+2020} which tend to show a lower rotation than the gas. 
However, the innermost RC within $\lesssim 0.5$ kpc is well connected to the RC in the CMZ and the derived mass distribution agrees well with the surface photometry result of $4\mu$ m, as shown by the dynamics as a whole.

\begin{figure}
\begin{center}     
\includegraphics[width=\lw]{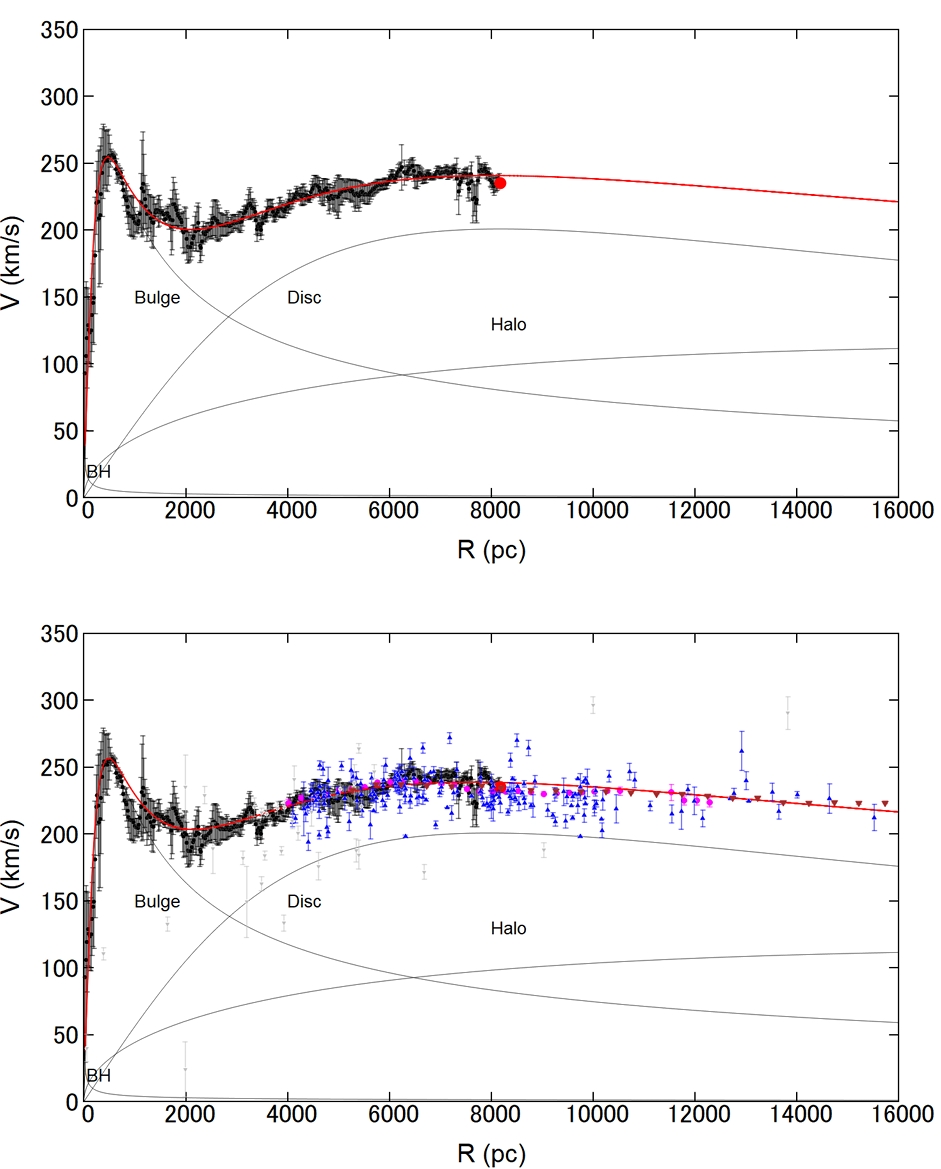} 
\end{center} 
\caption{{[Top] Fitting using HI+CO RC by TVM alone and [Bottom] using HI+CO + VERA + VLBA + GAIA RCs, showing almost equivalent fits.}
{Alt text: Fitting using the inner RC alone and using all RCs obtained in this paper.}
} 
\label{compare2}
\end{figure}

{In order to demonstrate how accurate is the inner RC determined by the VTM from the HI and CO lines alone, we compare the decomposition result with that using all the rotation velocities including VERA, VLBA, and Gaia.
Figure \ref{compare2} shows the comparison and we find that the two fit results agree with each other.}

\ss{Model fitting: decomposition into mass components} 
\label{rcfit}

We try to fit the RC by a simple model represented by superposition of the central black hole, bulge, disc, and dark halo as follows.
The central black hole is represented by a point mass of $4\times 10^6\Msun$,
\be V_1=131.5 ~{1 \over \sqrt{x}} ~\ekms ~\left(x={R\over 1~ \epc}\right). \ee 
The bulge and disc potentials are represented by the Plummer potential.
Since we treat the rotation velocity in the Galacic plane, or $z=0$, it is equivalent to the Miyamoto-Nagai (MN) potential \citep{mn1975} where the scale radius and height $R$ and $z$ ($a_r$ and $b_z$) are degenerate to
$a_i=\sqrt{a_{r,i}^2+b_{z,i}^2}$.
The circular velocities of the bulge and disc are given by
\be   V_i=V_{\rm i,0} ~\sqrt{-x~  {d\Phi_i \over dx}}~ \ekms ~\left(x={R\over a_i }\right),\ee 
where 
\be \Phi_i = {1\over \sqrt{1+x^2}},\ee
and $a_2=a_{\rm bulge}$ and $a_3=a_{\rm disc}$.
The total masses of the bulge and disc are calculated by 
\be M_i=a_i V_{i,0}^2/G.
\ee

The dark halo is assumed to have the NFW density profile, 
\be
\rho=\rho_0/[x(1+x)^2],
\ee 
with $x=R/h$.
The circular velocity at radius $R$ is given by
\be    V_4=V_{\rm h}~\sqrt{g(x) \over x} \ekms~\left(x={R\over h}\right),\ee
where 
\be g(x)=\int_0^x  {4 \pi \xi^2 d\xi\over \xi(1+\xi)^2},\ee
and  $\rho_0=V_{\rm h}^2/(G h^2)$.
The mass of DM within $R$, which is divergent, is given by
\be
M_{\rm DM}={4\pi h V_{\rm h}^2\over G}  \left[{\rm ln}\left(1+{R\over h}\right)+{1\over 1+R/h}-1\right].
\ee
The rotation velocity is calculated by      
\be      V=\sqrt{V_1^2+V_2^2+V_3^2+V_4^2}.\ee     

We comment that the mass distribution creating the Plummer-type potential has no singularity at the nucleus, whereas exponential type density distribution, including the de Vaucouleurs law, often assumed in the Galactic dynamics, yields a singular behavior of mass distribution at the nucleus. 
Moreover, the rotation curve beyond the scale radii is almost the same for the Plummer and exponential models. 
So, in this paper, we adopt the Plummer type.

By the least-squares fitting to the plot in figure \ref{compare}, we obtain the parameters listed in table \ref{tab_fit}.
The rotation curves calculated using these parameters are drawn in figure \ref{compare} superposed on the data.

\begin{table*}[]
    \caption{Decomposition of the rotation curve into bulge, disc, and dark halo.\\
    }
    \centering
    \begin{tabular}{cccc}
    \hline
    \hline
Component&          $ V_i$ & $a_i$& $M_i$\\
&         (\kms)&(pc)&($10^{11}\Msun$) \\
\hline 
Bulge &        406.2 $ \pm $          1.7&       332.8$ \pm $         3.7 &       0.127$\pm$       0.002\\
 Disc&        322.4 $ \pm $          0.6&      5624.8$ \pm $        46.2 &       1.352$\pm$       0.011\\
 Halo&         64.4 $ \pm $          0.4&     22379.1$ \pm $       684.2 &       0.599$\pm$       0.019\\
 \hline
 DMD at Sun $\dagger$\\
  (g cm$^{-3}$)&&&   $0.191\pm    0.006\times 10^{-24}$\\
 (GeV cm$^{-3}$) &&&       0.107 $\pm$        0.003\\
\hline 
    \end{tabular}\\
$\dagger$ Dark halo component alone fitted by the NFW profile. Contribution by the disk is not included.

    \label{tab_fit} 
\end{table*} 

\ss{{Dark matter fraction}}

\begin{figure}
\begin{center}     
\includegraphics[width=\lw]{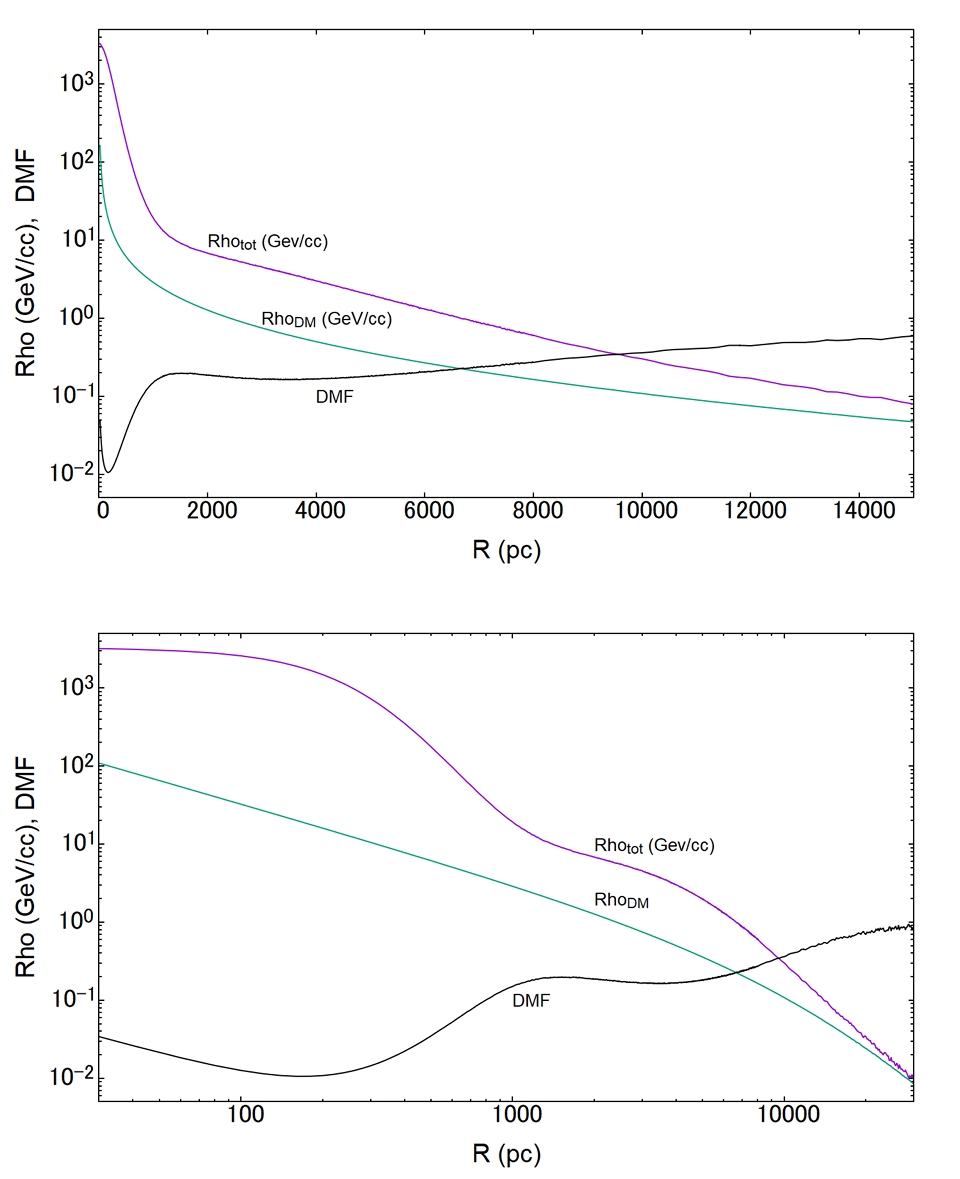} 
\end{center} 
\caption{{Total and DM mass densities and dark matter fraction, DMF $f_{\rm DM}=\rho_{\rm DM}/\rho_{\rm Tot}$, inside $R$, calculated using the parameters from the RC decomposition.}
{Alt text: Plot of the dark matter fraction.}
} 
\label{dmfrac}
\end{figure}

Using the decomposition of the RC, we can calculate the dark matter fraction (DMF), which is defined by the ratio of the dark matter density (DMD) to the total mass density, 
\be
f_{\rm DM}={\rho_{\rm DM}(R)\over \rho_{\rm Total}(R)}.
\label{eq_dmfrac}
\ee
Figure \ref{dmfrac} shows the calculated DMF using the fitted functions together with the total and DM densities.

The Galaxy is disc-dominant in the analyzed region at $R\lesssim 25$ kpc.
Within the solar circle, $R\lesssim8.2$ kpc, the DM mass shares DMF=0.1 of the total mass.
The disc remains dominant until $R=2$ kpc, within which the bulge becomes dominant.
The DMF decreases monotonically toward the center and becomes less than 0.03 inside the bulge.
However, DM is not negligible in any radius in the analyzed area and overcomes the disc within $\sim 1$ kpc, where DMF remains a few percent of the bulge.
Then, in the innermost region within $\sim 100$ pc, the DMF decreases again due to the black hole and the circumnuclear stellar core (not resolved here) despite the increase of the density toward the center as $\propto R^{-1}$ obeying the NFW function.

In the solar vicinity at $R=8.2$ kpc, the local DMD is estimated to be $\rho_{\rm DM}^\odot  =0.107 \pm        0.003$ Gev cm$^{-3}$ using the NFW function with the parameters listed in table \ref{tab_fit}.
{This value is much smaller than the current values of $\sim 0.3$ GeV cm$^{-3}$ estimated from various methods \citep{sofue2020}.
The reason for a lower value here is the new RC adopted in this paper, which is monotonically declining beyond the solar circle until the end of the fitting at $R\sim 25$ kpc.
However, the DMD contained by the disk component is not separated here, so the calculated DMD value for the NFW halo component should be taken as the lower limit of true DMD.
}

\ss{East-west (E/W) asymmetry}
\label{ss_ew}

\begin{figure}
\begin{center}  
\includegraphics[width=\lw]{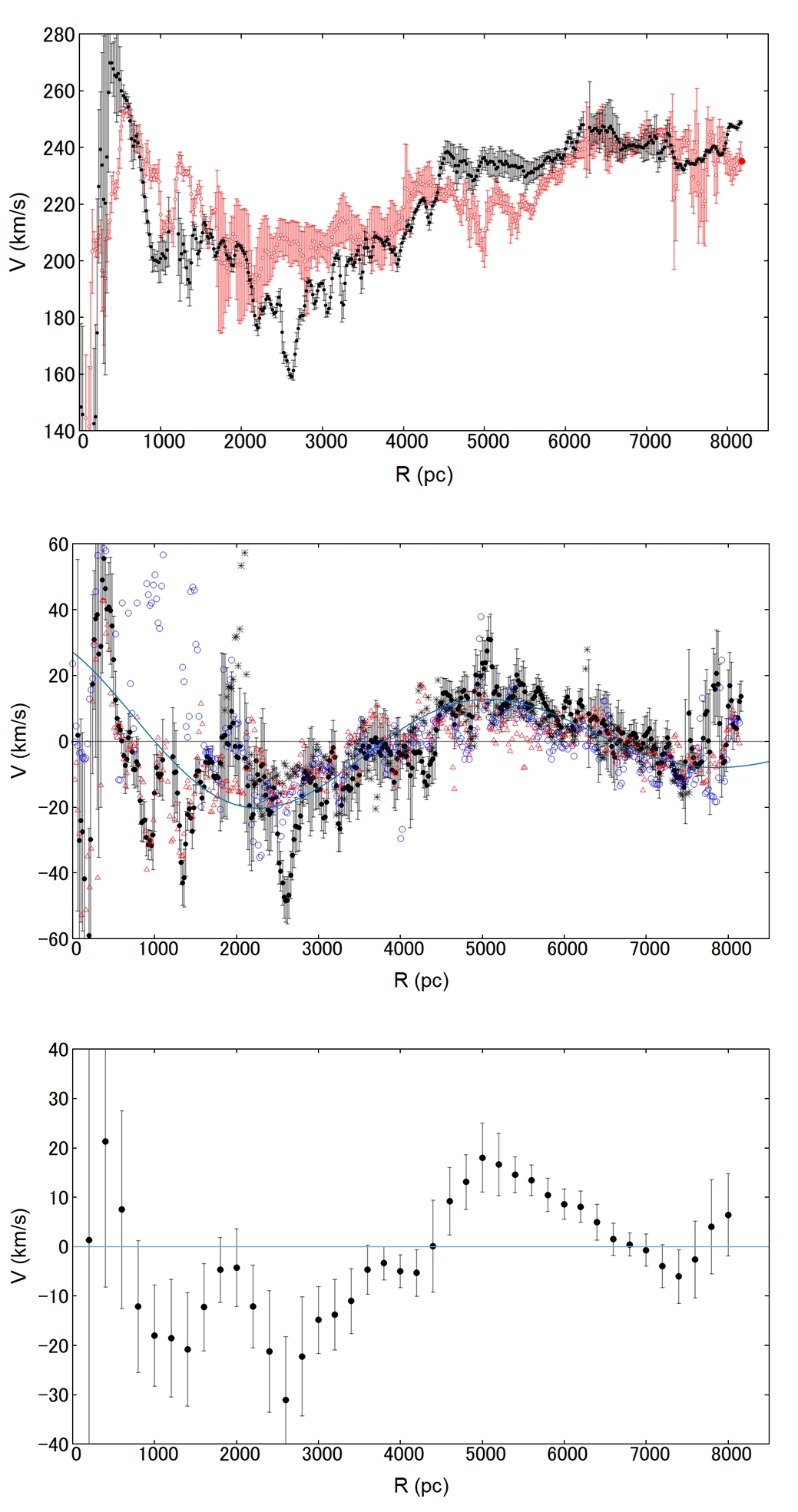}    
\end{center} 
\caption{[Top] Eastern (dot) and western (circle) RCs.
[Middle] East-west difference. Dots=all RC; circles=CO line RC from CfA; triangles=HI from HI4PI.
The curve represents $\delta \Vrot$ given by equation \ref{difsin}.
Note also that the variation is more sawtooth like with negative and positive peaks at $R=2.5$, 5 and 7.4 kpc.
{[Bottom] Same, but running average with a radial bin size of 200 pc.}
{Alt text: East/west asymmetry of the rotation curves.}
} 
\label{dif}
\end{figure}

A more detailed inspection of the inner RC in the east and west of the GC shown in figure \ref{bulge} reveals an east/west (E/W) asymmetry.
In figure \ref{dif} we plot the E and W rotation curves separately, and their difference, $\delta \vrot$, in the bottom panel. 
Black dots show the values using all data, open circles show those for the CfA-Chile CO data, triangles are for HI, and asterisques are for CO data from the Nobeyama 45-m and Mopra CO surveys data.
The asymmetry is visible in the independent observations and in the HI and CO lines, indicating that the sinusoidal variation is a dynamical effect, not dependent on the observed species.

The E/W asymmetry of RC can be explained by the non-circular motion induced by a bar potential \citep{1975PASJ...27...35M}. 
The high accuracy RC from the present analysis makes it possible to quantify the discussion with greater precision, including a new dynamical aspect, such as the sawtooth variation of RC.
The E/W difference may be approximately fitted by a sinusoidal curve with decreasing amplitude as follows as inserted in the plot: 
\be
\delta \Vrot
\sim ~45~ {\exp}\left({-{R\over3.5\ekpc}} \right) {\sin}\left(2 \pi {R-4\ekpc\over4.4\ekpc}\right) \ekms,
\label{difsin}
\ee 
The fact that the amplitude decreases with the radius indicates that the E/W asymmetry is caused by an internal mechanism such as a bar rather than an external origin such as the tide of the companion galaxy.  

In the mid-disc region at $R\sim 2$ to 6 kpc, where the sinusoidal variation appears most clearly, the amplitude is measured to be $\delta \vrot \simeq 15$ \kms.
This value may be used to estimate the bar potential by 
\be
{\partial \delta \Phi \over \partial R} \sim {\delta V^2 \over R},
\ee
which can be read as
\be
{GM_{\rm bar}\over R_y} \sim \delta \vrot^2,
\ee
where $R_y$ is the minor axial length of the bar (major axis length is $R_x$).
If we assume $R_y\sim 2$ and $R_x \sim 4$ kpc, we obtain a rough estimate of the mass of the bar at
$M_{\rm bar}\sim R_x \delta \vrot^2/G\sim 10^8\Msun$ for $\delta \vrot\sim 15$ \kms.
This is two orders of magnitude smaller than the disc ($\sim$total) mass inside $R\sim R_x\sim 4 \ekpc$,
$M_{\rm disc}(\le 4 \ekpc)\sim 4\times 10^{10}\Msun$, and the mass of the bulge, $M_{\rm bulge}(\le 4 \ekpc)\sim 10^{10}\Msun$.
However, note that the above estimate applies if the axial ratio of the bar is sufficiently large, here two.
If the bar is rounder, a higher mass is needed to cause the observed E/W asymmetry.

On the other hand, in the more inner region at $R\sim 2000$ pc, the $\delta \vrot$ plot is largely scattered due to both individual large errors and irregular variation of shorter wavelength and large amplitude of $|\delta \vrot|\sim 30$ to 50 \kms.
The large amplitude can be attributed to noncircular high-velocity motion inside the bar.
However, the irregular variations with shorter wavelengths, which are revealed in the present detailed analysis, cannot be attributed to a bar model, raising a new problem about the RC analysis. 

Another important aspect is the finer structure of the E/W variation.
Although the sinusoidal fit is reasonable to approximate the variation, the plot in the lower panel of figure \ref{dif} is more sawtooth-like than sinusoidal.
The plot has a clear negative peak at $R=2500$, a positive peak at 5000, and a negative peak at 7400 pc.
The plot is almost straight with a slope of
$d\delta V/dR  \sim \pm 12$ \kms kpc$^{-1}$ with respect to $R$ between the peaks.

The discontinuous turn of the gradient at these particular radii suggests that the $\delta \vrot$ variation is due to a discontinuity in the local flow velocity of the HI and CO gas.
Such a phenomenon is expected at galactic shock waves likely associated with the spiral arm and/or a bar.
However, the discontinuous property may not be due to a resonance phenomenon of the orbits caused by the long-range gravitational force of the galactic scale.

\ss{The bulge: On the axisymmetry assumption for the RC study}

For studying the mass distribution using the rotation curve, we assume that the galaxy is axisymmetric \citep{Sofue2017}.
Insofar as the disturbance is small, first-order analysis provides some meaningful information on the perturbed disc, such as a weak bar and spiral arms \citep{1975PASJ...27...35M,2021PASJ...73L..19S}

\begin{figure}
\begin{center}      
\includegraphics[width=\lw]{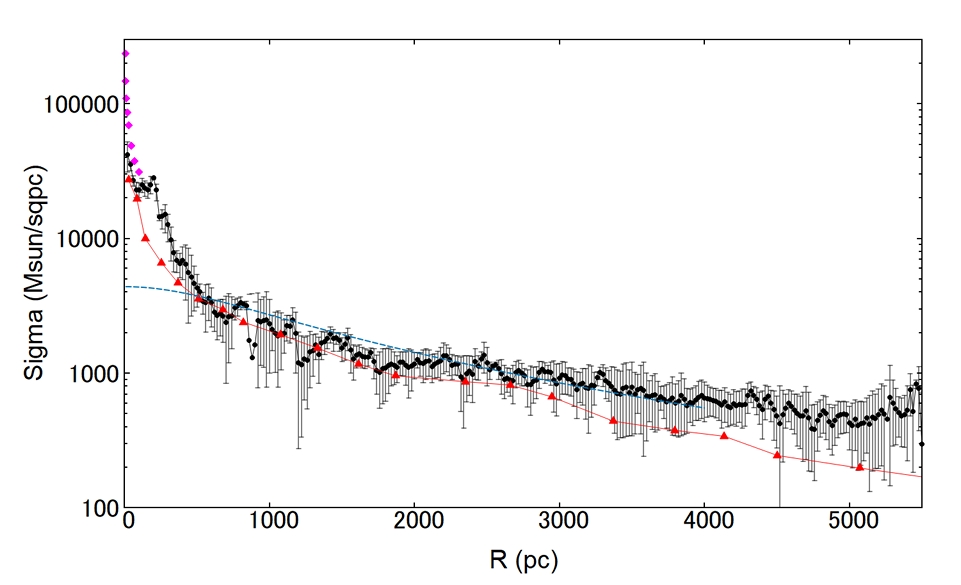}   
\end{center} 
\caption{
{Surface mass density (SMD) of the inner Milky Way in a semilogarithmic scaling.
Dots: SMD derived for the RM/TVM using equation \ref{eq_sigA} for a spherical assumption.
Diamonds: Observed SMD by infrared photometry of the GC \citep{2014A&A...566A..47S}.
Triangles: Observed SMD by infrared photometry of the bulge (projection onto the galactic plane) \citep{2017MNRAS.465.1621P}.
Dashed line: SMD calculated for the 'true RC' by simulation \citep{2015A&A...578A..14C}.}
{Alt text: Radial distributions of the surface density by photometric observations, using the present RC, and using an RC of a simulated galaxy. }
} 
\label{sigsim}
\end{figure}

In figure \ref{sigsim} we show the radial profile of the surface mass density calculated for the RC/TVM using equation \ref{eq_sigA} in the inner Milky Way by semilogarithmic scaling.
The radial profile derived here using the RC/TVM agrees with the photometric measurements of the mass distribution of the bulge \citep{2016A&A...587L...6V,2017MNRAS.465.1621P} and the Galactic center \citep{2014A&A...566A..47S}.
The mass of the bulge of $\sim 1-2\times 10^{10}\Msun$ obtained by photometry \citep{2016A&A...587L...6V, 2017MNRAS.465.1621P} also agrees with the present estimation of $10^{10}\Msun$.
So, the extence of a massive bulge in the Milky Way is evident.
This is consistent with the bulge + disc structure inferred from the photometric decomposition of large galaxy samples into the bulge + disc structure \citep{2011ApJS..196...11S,2012MNRAS.421.2277L}. 

Kinematically, most spiral galaxies have RCs that rise steeply in the central $\lesssim 1$ kpc \citep{1999ApJ...523..136S,1997PASJ...49...17S}, which may be attributed to a massive bulge or a noncircular flow parallel to the major axis of a bar. 
However, the probability of seeing a bar end-on is smaller than seeing it at a large angle at which the flow yields rather a lower velocity.
Therefore, the generally observed steep increase in RC may not be attributed to the bar but to a fundamental mass structure of the bulge.

\ss{On the simulation approach}
\label{onthesimulation}

If a non-axisymmetric potential, such as due to a bar, is deep, it creates a noncircular motion and hinders the inherent RC.
So, careful attention is needed when converting the RC to mass in such a case.
If the assumption of axisymmetry is removed and an RC still has to be used to discuss the dynamics of a galaxy, we need to solve the inverse problem from the one-dimensional plot of $\Vrot(R)$ to a two-dimensional potential (or three).
But this is mathematically impossible.

Attempts have been made to solve this problem using a numerical simulation by \citet{2015A&A...578A..14C}.
They showed that the central high-velocity peak of the RC/VTM can be produced by a noncircular flow because of the presence of a bar.
However, their 'true RC' of the underlying model 'Milky Way' has a rigid-body rotation linearly increasing from zero near the center with a slope of $\sim 250$ \kms kpc$^{-1}$ (see their Fig. 4), indicating that the stellar density is nearly constant and the SMD is flat near the center.
In figure \ref{sigsim} we show the SMD calculated for the 'true RC', which clearly contradicts the observed SMD by infrared photometry.
Because such a bulgeless model does not simulate the Milky Way, their result may not be compared with the current RCs obtained for the MW including the result here.
However, the simulation might be useful for a particular type of disc galaxy, such as dwarf galaxies and galaxies with low surface brightness that show a mild rise in RC in the center.

\section{Summary}
\label{summary}

We derived the inner rotation curve of the Milky Way by applying the terminal velocity method to the longitude-velocity diagrams made from the large-scale survey data of the Galactic plane in the HI and \co\ lines.
We confirmed that the RC agrees well with the RCs derived from the astrometric measurements of stars and maser sources and combined them to construct a unified RC up to $R\sim 25$ kpc.
A detailed comparison of the eastern ($l\ge 0\deg$) and western ($< 0\deg$) RCs in the HI and CO lines within the solar circle ($R\le 8.2$ kpc) allowed the creation of an accurate E/W asymmetry curve of the velocity difference.
We showed that the E/W assymetry is fitted by a sinusoidal function of the radius with the amplitude decreasing with the distance from the Galactic Center.

\section*{Acknowledgments}
Data analysis was performed at the NAOJ Astronomy Data Center. 
The authors thank Prof. Tomoya Hirota of the NAOJ/VERA for providing the machine-readable table of the VERA RC.
The Nobeyama 45 m radio telescope is operated by Nobeyama Radio Observatory, a branch of the National Astronomical Observatory of Japan. 
The FUGIN CO data were retrieved from the JVO portal (url {http://jvo.nao.ac.jp/portal}) operated by ADC/NAOJ.
The authors thank Dr. Shinji Fujita of the Institute of Statistical Mathematics for useful comments on data analysis of the fits-cube data.
They thank the FUGIN project members for the CO survey data observed by the Nobeyama 45-m telescope,
Prof. Tomoharu Oka and the radio astronomy group of Keio University for the archival GC survey data using the Nobeyama 45 m telescope,
Dr. Michael Burton of the University of New South Wales and the Armagh Observatory and Planetarium for the archival CO survey data with Mopra, 
and Dr. Graeme Wong for kindly supporting remote observations from Nagoya University.
The Mopra radio telescope is part of the Australia Telescope National Facility, which is funded by the Australian Government for operation as a National Facility managed by CSIRO. The University of New South Wales Digital Filter Bank used for the observations with the Mopra Telescope was provided with support from the Australian Research Council.
{We thank the anonymous referee for the valuable comments to improve the paper.}

\section*{Conflict of interests}
The authors declare that there are no conflicts of interest.

\section*{Data availability}
The obtained rotation curve is listed in the tables in the Appendix.
The survey data used in this article are available electronically, as described in the text in appropriate sections.

\begin{appendix}

\section{Averaged rotation curves and tables}

\begin{figure}
\begin{center}    
\includegraphics[width=\lw]{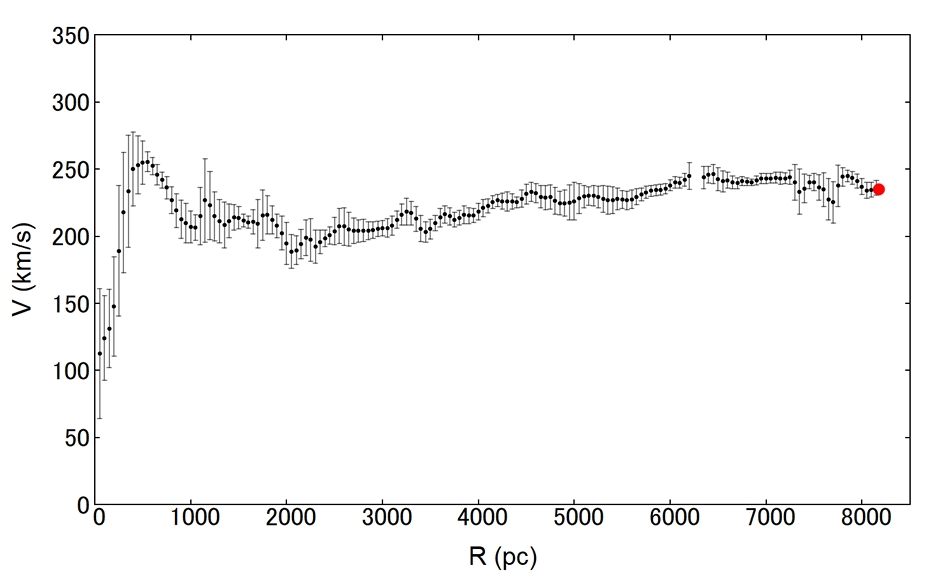}   
\end{center} 
\caption{Inner rotation curve plotted every 50 pc.
{Alt text: Rotation curve with 50 pc radial increment.}}
\label{rc50}
\begin{center}        
\includegraphics[width=\lw]{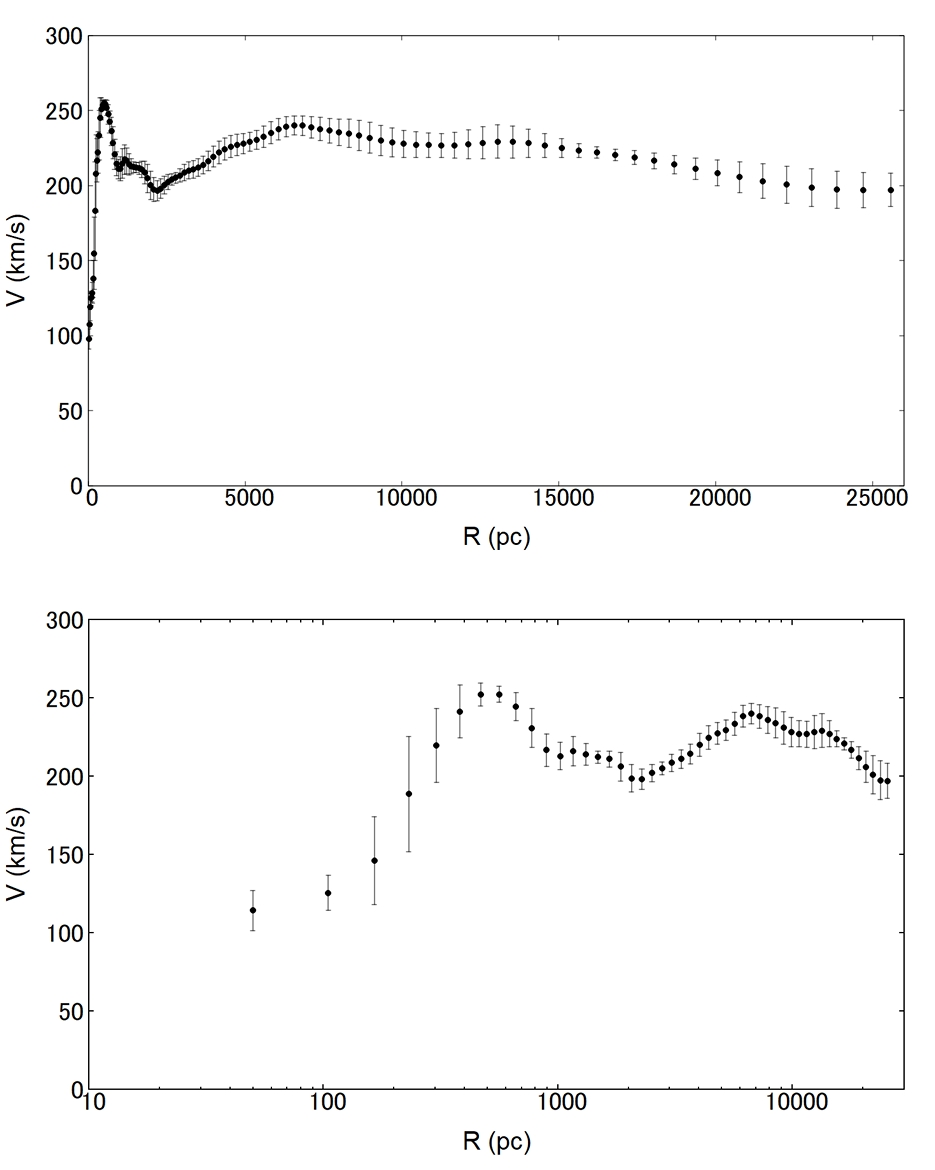}  
\end{center}   
\caption{{
Running averaged unified RC combined with the VERA, VLBA and Gaia DR2 results with increasing radius increment. 
{Alt text: Unified rotation curve of the Milky Way combined with the astrometric results.} 
} }
\label{urcga}
\label{urc}
\end{figure}

In this Appendix we present running averaged rotation curves. Figure \ref{rc50} shows the inner RC with a bin size of $\delta R=50$ pc, and figure \ref{urc} is a unified RC combined with the VERA and VLBA results with increasing radius increment.
The plotted values are listed in tables \ref{tab_rc50} and \ref{taburc}, respectively.

\begin{table}[]
    \caption{Table of the inner rotation curve for radius increment of $\delta R=50$ pc. Same as figure \ref{rc50}}
    \centering
    \begin{tabular}{ccc}
    \hline
    \hline
$R$ &  $\vrot$ &  $\delta \vrot$   \\
(pc) &   (\kms)& (\kms) \\
\hline
       50.0 &     112.37 &      48.44 \\
      100.0 &     124.08 &      31.52 \\
      150.0 &     131.24 &      29.01 \\
      200.0 &     147.71 &      36.91 \\
      250.0 &     189.03 &      48.59 \\
      300.0 &     217.69 &      44.75 \\
      350.0 &     233.51 &      41.81 \\
      400.0 &     250.16 &      27.45 \\
      450.0 &     253.10 &      21.51 \\
      500.0 &     254.84 &      15.97 \\
      550.0 &     255.43 &       7.42 \\
      600.0 &     252.31 &       6.31 \\
      650.0 &     245.68 &       7.57 \\
      700.0 &     241.91 &       5.90 \\
      750.0 &     236.11 &       8.09 \\
      800.0 &     226.87 &      10.02 \\
      850.0 &     219.05 &      12.76 \\
      900.0 &     212.54 &      14.19 \\
      950.0 &     209.80 &      14.52 \\
     1000.0 &     207.01 &      11.73 \\
     1050.0 &     206.50 &       9.60 \\
     1100.0 &     214.87 &      21.39 \\
     1150.0 &     226.65 &      31.04 \\
     1200.0 &     222.92 &      25.42 \\
     1250.0 &     214.84 &      18.23 \\
     1300.0 &     211.16 &      16.17 \\
     1350.0 &     208.29 &      16.99 \\
     1400.0 &     211.41 &      12.76 \\
     1450.0 &     214.19 &       9.56 \\
     1500.0 &     213.71 &       8.36 \\
     1550.0 &     211.72 &       4.83 \\
     1600.0 &     210.44 &       4.59 \\
     1650.0 &     210.61 &       8.98 \\
     1700.0 &     209.45 &      17.95 \\
     1750.0 &     215.64 &      18.73 \\
     1800.0 &     215.99 &      14.32 \\
     1850.0 &     211.93 &      10.44 \\
     1900.0 &     207.65 &       8.67 \\
     1950.0 &     202.36 &      12.73 \\
     2000.0 &     194.58 &      15.60 \\
     2050.0 &     188.62 &      12.38 \\
     2100.0 &     189.45 &      10.66 \\
     2150.0 &     194.16 &      10.80 \\
     2200.0 &     198.84 &      13.45 \\
     2250.0 &     197.20 &      15.81 \\
     2300.0 &     192.24 &      12.39 \\
     2350.0 &     195.77 &       8.77 \\
     2400.0 &     198.47 &       6.27 \\
     2450.0 &     200.52 &       6.61 \\
     2500.0 &     203.73 &      10.17 \\
     2550.0 &     207.47 &      13.08 \\
     2600.0 &     207.31 &      13.99 \\
     2650.0 &     205.15 &      12.51 \\
     2700.0 &     204.03 &       9.48 \\
     2750.0 &     203.90 &       8.18 \\
     2800.0 &     204.22 &       8.17 \\

\hline
    \end{tabular} 
    \label{tab_rc50}
\end{table}

\setcounter{table}{2}
\begin{table}
    \centering
    \begin{tabular}{ccc}
    \hline
    \hline
$R$ &  $\vrot$ &  $ \delta \vrot$   \\
(pc) &   (\kms)& (\kms)   \\
\hline
     2850.0 &     203.88 &       6.65 \\
     2900.0 &     204.47 &       6.07 \\
     2950.0 &     205.66 &       5.66 \\
     3000.0 &     206.03 &       5.60 \\
     3050.0 &     206.17 &       6.95 \\
     3100.0 &     208.00 &       7.20 \\
     3150.0 &     212.23 &       6.33 \\
     3200.0 &     216.11 &       7.78 \\
     3250.0 &     218.39 &       9.95 \\
     3300.0 &     217.16 &       8.75 \\
     3350.0 &     212.89 &       9.34 \\
     3400.0 &     205.52 &       9.74 \\
     3450.0 &     203.11 &       7.69 \\
     3500.0 &     205.36 &       7.47 \\
     3550.0 &     209.77 &       5.73 \\
     3600.0 &     213.80 &       6.68 \\
     3650.0 &     216.21 &       6.31 \\
     3700.0 &     214.96 &       6.41 \\
     3750.0 &     212.31 &       5.37 \\
     3800.0 &     213.77 &       5.33 \\
     3850.0 &     215.93 &       6.42 \\
     3900.0 &     215.36 &       5.83 \\
     3950.0 &     215.55 &       5.17 \\
     4000.0 &     218.39 &       6.30 \\
     4050.0 &     221.01 &       6.42 \\
     4100.0 &     222.62 &       5.05 \\
     4150.0 &     225.36 &       4.57 \\
     4200.0 &     226.70 &       4.47 \\
     4250.0 &     226.11 &       5.74 \\
     4300.0 &     226.01 &       7.04 \\
     4350.0 &     225.83 &       5.72 \\
     4400.0 &     225.19 &       4.15 \\
     4450.0 &     228.02 &       6.23 \\
     4500.0 &     231.43 &       7.24 \\
     4550.0 &     233.14 &       7.17 \\
     4600.0 &     232.16 &       7.20 \\
     4650.0 &     229.42 &       8.19 \\
     4700.0 &     228.72 &       9.17 \\
     4750.0 &     229.26 &       8.91 \\
     4800.0 &     226.39 &       9.83 \\
     4850.0 &     224.64 &       8.93 \\
     4900.0 &     224.64 &       9.69 \\
     4950.0 &     225.06 &      12.95 \\
     5000.0 &     226.01 &      14.09 \\
     5050.0 &     228.09 &      11.05 \\
     5100.0 &     229.46 &       8.39 \\
     5150.0 &     230.03 &       6.97 \\
     5200.0 &     230.01 &       7.53 \\
     5250.0 &     229.24 &       8.23 \\
     5300.0 &     227.69 &       9.84 \\
     5350.0 &     226.84 &      10.46 \\
     5400.0 &     226.82 &      10.02 \\
     5450.0 &     227.78 &       7.97 \\
     5500.0 &     227.22 &       6.76 \\
     5550.0 &     226.70 &       6.96 \\
     5600.0 &     227.43 &       6.96 \\

\hline
    \end{tabular} 
\end{table}

\setcounter{table}{2}
\begin{table}[]
    \centering
    \begin{tabular}{ccc}
    \hline
    \hline
$R$ &  $\vrot$ &  $ \delta \vrot$   \\
(pc) &   (\kms)& (\kms)   \\
\hline
     5650.0 &     229.31 &       5.85 \\
     5700.0 &     231.14 &       4.86 \\
     5750.0 &     232.77 &       4.60 \\
     5800.0 &     233.86 &       4.09 \\
     5850.0 &     234.24 &       3.84 \\
     5900.0 &     234.56 &       3.97 \\
     5950.0 &     235.23 &       3.81 \\
     6000.0 &     237.94 &       4.16 \\
     6050.0 &     240.01 &       4.17 \\
     6100.0 &     239.73 &       4.00 \\
     6150.0 &     241.85 &       4.75 \\
     6200.0 &     244.97 &       9.92 \\
     6350.0 &     243.87 &       8.28 \\
     6400.0 &     245.95 &       6.15 \\
     6450.0 &     246.23 &       7.10 \\
     6500.0 &     242.29 &       8.55 \\
     6550.0 &     240.85 &       7.88 \\
     6600.0 &     241.63 &       5.88 \\
     6650.0 &     240.17 &       4.80 \\
     6700.0 &     239.48 &       3.86 \\
     6750.0 &     240.86 &       3.32 \\
     6800.0 &     240.59 &       2.70 \\
     6850.0 &     239.89 &       2.37 \\
     6900.0 &     241.49 &       3.24 \\
     6950.0 &     243.08 &       3.82 \\
     7000.0 &     242.90 &       3.86 \\
     7050.0 &     243.14 &       3.62 \\
     7100.0 &     243.58 &       4.11 \\
     7150.0 &     243.20 &       4.48 \\
     7200.0 &     243.14 &       3.92 \\
     7250.0 &     243.96 &       5.07 \\
     7300.0 &     240.29 &      13.24 \\
     7350.0 &     233.03 &      16.85 \\
     7400.0 &     235.54 &      10.65 \\
     7450.0 &     240.19 &       5.00 \\
     7500.0 &     240.34 &       6.61 \\
     7550.0 &     236.22 &       9.59 \\
     7600.0 &     234.68 &      12.71 \\
     7650.0 &     227.46 &      15.29 \\
     7700.0 &     225.35 &      15.36 \\
     7750.0 &     237.62 &      15.29 \\
     7800.0 &     244.25 &       6.90 \\
     7850.0 &     244.75 &       4.19 \\
     7900.0 &     243.44 &       4.71 \\
     7950.0 &     241.11 &       5.15 \\
     8000.0 &     236.95 &       5.91 \\
     8050.0 &     234.18 &       5.95 \\
     8100.0 &     234.46 &       5.46 \\
     8150.0 &     236.12 &       5.50 \\

\hline
Sun: 8178.0 & 235.1 & --- \\
\hline
    \end{tabular} 
\end{table}


\begin{table}[]
    \caption{{Running averaged rotation curve combined with the VERA, VLBA and Gaia DR2 results with increasing radius increment. Same as figure \ref{urcga}.}}
    \centering
    \begin{tabular}{ccc}
    \hline
    \hline
$R$ &  $\vrot$ &  $\delta \vrot$   \\
(pc) &   (\kms)& (\kms) \\
\hline
       20.000 &       97.632 &        6.707 \\
       41.000 &      107.342 &       10.051 \\
       63.037 &      119.041 &        9.226 \\
       86.151 &      124.910 &        4.900 \\
      110.381 &      125.371 &        3.043 \\
      135.769 &      128.470 &        6.887 \\
      162.357 &      137.813 &       12.394 \\
      190.190 &      154.880 &       23.963 \\
      219.312 &      183.077 &       32.886 \\
      249.773 &      207.928 &       25.669 \\
      281.619 &      216.720 &       14.441 \\
      314.901 &      221.879 &       13.829 \\
      349.671 &      233.510 &       16.615 \\
      385.983 &      245.147 &       13.116 \\
      423.892 &      251.093 &        7.216 \\
      463.455 &      253.395 &        3.462 \\
      504.732 &      254.547 &        2.095 \\
      547.782 &      254.285 &        3.009 \\
      592.670 &      251.706 &        5.146 \\
      639.460 &      247.383 &        6.303 \\
      688.219 &      242.342 &        7.298 \\
      739.016 &      236.111 &        9.279 \\
      791.923 &      228.492 &       10.591 \\
      847.013 &      220.799 &       10.044 \\
      904.363 &      214.608 &        8.057 \\
      964.051 &      211.062 &        6.438 \\
     1026.158 &      211.312 &        7.929 \\
     1090.768 &      214.694 &        9.921 \\
     1157.967 &      217.278 &        9.707 \\
     1227.844 &      216.503 &        8.609 \\
     1300.491 &      214.168 &        6.973 \\
     1376.004 &      212.728 &        5.008 \\
     1454.479 &      212.248 &        3.630 \\
     1536.018 &      211.839 &        3.316 \\
     1620.725 &      211.395 &        3.971 \\
     1708.707 &      210.753 &        5.378 \\
     1800.075 &      208.734 &        7.567 \\
     1894.944 &      204.698 &        9.391 \\
     1993.431 &      200.141 &        9.344 \\
     2095.658 &      197.227 &        7.891 \\
     2201.751 &      196.670 &        6.670 \\
     2311.838 &      197.982 &        6.294 \\
     2426.054 &      200.278 &        5.984 \\
     2544.536 &      202.481 &        5.055 \\
     2667.426 &      203.992 &        3.881 \\
     2794.869 &      205.184 &        3.564 \\
     2927.018 &      206.748 &        4.443 \\
     3064.027 &      208.629 &        5.363 \\
     3206.057 &      210.091 &        5.771 \\
     3353.274 &      210.953 &        5.841 \\
\hline
    \end{tabular} 
    \label{taburc} 
\end{table}

\setcounter{table}{2}
\begin{table}[]
    \centering
    \begin{tabular}{ccc}
    \hline
    \hline
$R$ &  $\vrot$ &  $\delta \vrot$   \\
(pc) &   (\kms)& (\kms) \\
\hline

     3505.848 &      211.937 &        5.810 \\
     3663.955 &      213.725 &        5.994 \\
     3827.776 &      216.347 &        6.493 \\
     3997.498 &      219.333 &        7.039 \\
     4173.313 &      222.109 &        7.375 \\
     4355.422 &      224.311 &        7.444 \\
     4544.027 &      225.903 &        7.320 \\
     4739.341 &      227.077 &        7.065 \\
     4941.580 &      228.056 &        6.692 \\
     5150.969 &      229.050 &        6.358 \\
     5367.738 &      230.416 &        6.446 \\
     5592.128 &      232.475 &        6.972 \\
     5824.381 &      235.036 &        7.334 \\
     6064.753 &      237.482 &        7.156 \\
     6313.501 &      239.252 &        6.652 \\
     6570.898 &      240.049 &        6.341 \\
     6837.218 &      239.847 &        6.567 \\
     7112.748 &      238.897 &        7.143 \\
     7397.781 &      237.654 &        7.714 \\
     7692.620 &      236.513 &        8.206 \\
     7997.577 &      235.576 &        8.738 \\
     8312.975 &      234.660 &        9.362 \\
     8639.144 &      233.464 &        9.963 \\
     8976.425 &      231.837 &       10.271 \\
     9325.171 &      230.056 &       10.090 \\
     9685.744 &      228.631 &        9.607 \\
    10058.518 &      227.749 &        9.115 \\
    10443.876 &      227.254 &        8.617 \\
    10842.216 &      226.922 &        8.064 \\
    11253.946 &      226.712 &        7.838 \\
    11679.485 &      226.864 &        8.465 \\
    12119.269 &      227.537 &        9.682 \\
    12573.740 &      228.497 &       10.734 \\
    13043.362 &      229.187 &       11.093 \\
    13528.606 &      229.101 &       10.635 \\
    14029.959 &      228.155 &        9.500 \\
    14547.926 &      226.652 &        7.911 \\
    15083.023 &      224.982 &        6.161 \\
    15635.780 &      223.402 &        4.646 \\
    16206.750 &      221.937 &        3.806 \\
    16796.496 &      220.422 &        3.841 \\
    17405.600 &      218.638 &        4.466 \\
    18034.660 &      216.458 &        5.297 \\
    18684.295 &      213.899 &        6.189 \\
    19355.139 &      211.112 &        7.243 \\
    20047.850 &      208.302 &        8.617 \\
    20763.098 &      205.587 &       10.192 \\
    21501.578 &      202.977 &       11.574 \\
    22264.008 &      200.539 &       12.401 \\
    23051.119 &      198.526 &       12.562 \\
    23863.670 &      197.253 &       12.227 \\
    24702.439 &      196.827 &       11.680 \\
    25568.236 &      197.072 &       11.103 \\

\hline
    \end{tabular} 
    \label{tab_urc}
\end{table}


 \end{appendix}
 
\end{document}